\documentclass[acmsmall,screen]{acmart}
\AtBeginDocument{%
  }

\usepackage{makecell}
\usepackage{soul}
\usepackage{xcolor} 
\sethlcolor{yellow!30} 

\setcopyright{acmlicensed}
\copyrightyear{2018}
\acmYear{2027}
\acmDOI{XXXXXXX.XXXXXXX}
\acmConference[GROUP '27]{The ACM international conference on
Supporting Group Work}{January 10--13,
  2027}{Georgia, USA}
\acmISBN{978-1-4503-XXXX-X/2018/06}

\begin{document}

\title{AMINA: The Inclusive and Accountable AI for Marginalized Immigrant Nonprofit Assistance}

\author{Maryam Mokhberi}
\email{maryam@cs.toronto.edu}
\orcid{0000-0001-6470-8647}
\affiliation{%
  \institution{University of Toronto}
  \city{Toronto}
  \state{Ontario}
  \country{Canada}
}

\author{Dipto Das}
\email{dd749@cornell.edu}
\orcid{0000-0001-5704-8804}
\affiliation{%
  \institution{Cornell University}
  \city{Ithaca}
  \state{NY}
  \country{USA}
}

\author{Syed Ishtiaque Ahmed}
\email{ishtiaque@cs.toronto.edu}
\orcid{0000-0003-2452-0687}
\affiliation{%
  \institution{University of Toronto}
  \city{Toronto}
  \state{Ontario}
  \country{Canada}
}

\renewcommand{\shortauthors}{Mokhberi et al.}

\begin{abstract}
Immigrant-led nonprofit groups, particularly those operating in politically sensitive contexts, face exclusion from formal registries and digital platforms. This paper reports a three-phase mixed-methods study with Iranian immigrant nonprofit practitioners: 27 semi-structured interviews, a co-design session, and 7 evaluation and feedback interviews on a prototyped AI assistant, AMINA. Our findings highlight how legitimacy barriers, capacity gaps, and politically charged misinformation constrain nonprofit operations. We translate these insights into design goals for an inclusive nonprofit AI assistant: support for everyday group operations, recognition of informal nonprofit efforts, proactive countering of misinformation, and multilingual, accessible interaction. User evaluations show AMINA’s potential to reduce reporting burdens and foster transparency through proactive reminders, and catalyze collaboration across dispersed networks. We contribute to CSCW and HCI by characterizing the cooperative work of transnational immigrant nonprofits, extending scholarship on informality and misinformation, and demonstrating how AI can act as a collaborative partner that strengthens, rather than displaces, the human connections at the core of nonprofit ecosystems, while also posing major risks.

\end{abstract}



\received{31 March 2026}
\received[revised]{31 March 2026}
\received[accepted]{31 March 2026}

\maketitle

\section{Introduction}\label{introduction}


In North America, the nonprofit sector~\cite{noauthor_public_2025} contributes roughly 5\% of GDP and accounts for a considerable share of employment \cite{noauthor_types_2025}. Over the years, computing technologies have become integrated into nonprofit workflows, with the most recent Artificial Intelligence (AI) advances presenting new opportunities for fundraising, operational management, and donor outreach in this sector. Yet these benefits are unevenly distributed as under-resourced and marginalized nonprofit groups face barriers to adopting AI tools. They may also experience unanticipated harms from systemic inequalities in the technology and data ecosystem ~\cite{bopp_disempowered_2017, klein_data_2024}. This study investigates the nonprofit groups led by immigrants of Iranian descent (IINGs) in the Global North, specifically those focused on charitable efforts. We assess their interactions with emergent AI technologies and the socio-technical constraints they face under the shadow of international politics and access barriers. We propose designing an inclusive, accountable AI assistant tailored to the operational needs and unique challenges of marginalized nonprofits in this context.

Within the human-computer interaction (HCI) and computer-supported cooperative work (CSCW) fields, scholars have examined nonprofit practices and public engagement with vulnerable communities through the design of civic technologies and community-based organizing, highlighting the socio-technical constraints that shape underserved groups~\cite{ chordia_social_2024, bopp_voices_2020, sackitey_virtual_2025, goecks_charitable_2008}. While this scholarship provides critical insights into how organizations navigate resource scarcity and structural exclusion, it has primarily focused on formal nonprofits and mutual aid groups operating in Western contexts, as well as on development-oriented organizations in the Global South. What remains under-examined is a distinct class of nonprofit efforts: immigrant-based nonprofit groups whose operations span borders and whose vulnerabilities stem from complex geopolitical constraints. Viewed through a CSCW lens, these groups are sites of distributed cooperative work in a demanding form: interdependent activities carried out by volunteers, donors, and in-country partners who are separated by distance, language, and legal regimes~\cite{olson_distance_2000}, and who hold their work together through continuous articulation work, the often-invisible labor of aligning tasks, actors, and resources~\cite{ackerman_intellectual_2000,star_layers_1999,suchman_supporting_1995,schmidt_taking_1992}. Understanding how such cooperative arrangements are accomplished, and how emerging AI reshapes them, is a question of direct concern to GROUP and CSCW scholarship.

For immigrants, group-based charitable work supports not just the needs of others but also preserves their own cultural identity, belonging, and transnational connections to their countries of origin~\cite{bhugra_migration_2004, merz_diasporas_2007,zhou_homeland_2016,vertovec_migrant_2006}. However, many of these efforts operate informally, outside official recognition by the State, due to unique challenges related to immigration status, cultural and linguistic marginalization, and transnational political tensions \cite{zhang_migrants_2023}. As a result, despite their importance, they are often excluded from formal nonprofit registration systems, funding opportunities, and technology platforms~\cite{yousefi_remittance_2025}.

Motivated by previous research on immigrants’ multifaceted lives entangled with socio-politics, including charitable efforts \cite{yousefi_remittance_2025, paula_d_johnson_diaspora_2007, baig_re_2016, desouza_transnational_2023}, this study investigates the charitable giving practices of Iranian immigrants who operate under complex international political conditions, their engagement with computing and AI technologies, and the challenges and opportunities that arise. The Iranian immigrant population, estimated at 4–5 million worldwide \cite{mostafavi_mobasher_iranian_2018,noauthor_iranian_2025}, reflects a history shaped by both opportunity and forced displacement. While many have immigrated for education and economic advancement, others have fled political repression, religious persecution, or systemic discrimination in their country. Today, large Iranian diaspora communities around the world maintain strong cultural ties and frequently direct charitable donations toward their home country, Iran. Yet these efforts face distinctive geopolitical constraints, including licensing delays, heightened scrutiny, and legal barriers rooted in decades of strained relations with Western governments. International sanctions on Iran have made banks and service providers adopt “de-risking” practices that block or delay even humanitarian transactions with Iran despite formal exemptions. The impact of these political tensions on the financial practices of Iranian immigrants is detailed in the work of Rohanifar et al.~\cite{rohanifar_money_2021}.

At the community level, these geopolitical pressures have fostered informal, small-scale initiatives that rely heavily on digital tools. Messaging platforms, particularly Telegram, have become critical infrastructures for coordination, offering accessible features such as seamless group chats, file sharing, and tiered permissions. Nearly every diaspora community or friend group sustains its own charitable subgroup through such platforms. These messaging groups function, in effect, as improvised coordination mechanisms and common information spaces~\cite{schmidt_coordination_1996, bannon_constructing_1997}: they hold shared records of beneficiaries and expenses, allocate tasks among volunteers, and mediate awareness of who is doing what across continents, providing coordination functions that designated nonprofit platforms offer only to formally registered organizations. While these practices keep grassroots aid active, they remain structurally invisible to the nonprofit support resources, which typically require formal nonprofit status. While this has been the general political and socio-technical context around IINGs over the last few decades, the recent US-Israel-Iran wars in June 2025 and March 2026 have intensified the challenges, from a significantly heightened number of beneficiaries in need of aid to blocked internet communication with Iran and precarious transnational communications, effects that are not only important but also time-sensitive.

In this context, we aim to answer the following research questions guided by CSCW principles:

\begin{quote}
    
    RQ1: How do Iranian immigrant nonprofit groups coordinate charitable work across formal and informal arrangements, and how do existing digital infrastructures play a role in their interaction?
    
    RQ2:How can we leverage design principles to support the cooperative work of IINGs in harnessing emerging AI technologies while mitigating their adverse effects? 

\end{quote}

We conducted a three-phase mixed-methods study to investigate the RQs, comprising semi-structured interviews (phase 1), a co-design session (phase 2), and evaluation interviews with a high-fidelity prototype (phase 3). 
In phase 1, semi-structured interviews with 27 IING practitioners revealed persistent technological exclusion due to complex digital infrastructures and a lack of formal nonprofit status, alongside reputational harm caused by misinformation and geopolitical bias; these findings pointed to the need for digital tools that support routine nonprofit operations while being sensitive to political distrust and platform exclusion. 
These insights informed four design goals: (1) supporting everyday nonprofit tasks, (2) recognizing informal and unregistered efforts, (3) enabling proactive credibility-building and misinformation response, and (4) ensuring accessibility across diverse languages and levels of digital literacy.
In phase 2, guided by principles of design justice \cite{costanza-chock_design_2020}, which argues systems should be designed by those most affected by them, participants engaged in co-design activities to articulate expectations, develop user personas, and propose concrete design elements, which refined these goals into actionable features and directly shaped the design of an AI for Marginalized Immigrant Nonprofit Assistance, AMINA.
Finally, in phase 3, we developed a high-fidelity prototype of AMINA and conducted evaluation interviews with seven practitioners, which demonstrated how these design elements could be operationalized in practice and surfaced further refinements, including the need for alternative recognition mechanisms for informal groups, careful prebunking of misinformation, bridging capacity gaps, and broader design directions such as volunteer-in-the-loop workflows, proactivity, and inter-group collaboration.

Overall, AMINA is a contextually grounded AI assistant that supports routine nonprofit operations
while fostering inclusion, trust, collaboration, and organizational sustainability. This work makes an empirical contribution to CSCW by unpacking the cooperative work of transnational nonprofit groups operating under political tension, including the division of labor, coordination practices, and articulation work through which dispersed practitioners sustain collective charitable action, and a design contribution by showing how an AI assistant can be positioned as a collaborative partner that augments, rather than substitutes for, this cooperative fabric.

\section{Related Work}

Immigrants constitute an essential part of Global North societies, adding cultural diversity while sustaining strong ties to their countries of origin \cite{noauthor_immigrants_2022, noauthor_migration_2024}. Many contribute to transnational nonprofit initiatives that support underserved populations abroad \cite{merz_diasporas_2007, vertovec_migrant_2006, zhou_homeland_2016}. Although HCI and related fields have examined immigrant experiences and nonprofit technologies separately, the cooperative work at their intersection remains underexplored: namely, how immigrant collectives organize, coordinate, and sustain transnational charitable action, and how computing technologies and AI participate in that cooperation.

\subsection{Immigrant Identity and Transnational Nonprofit Practice}
\label{immigrant identity}
The lives of immigrants, especially in the early years post-immigration, are impacted by psychological distress and social challenges resulting from the loss of social support systems, cultural dislocation, and the challenges of adapting to a new environment. This process, known as acculturation, often leads to immigrants feeling marginalized and separated from the host society. However, many immigrants find it beneficial to integrate into the host society by adopting certain cultural traits of the new country while maintaining their original cultural identity. Seeking social support, engaging in cultural activities, and taking a proactive approach to overcoming immigration challenges are healthy ways of acculturation \cite{bhugra_migration_2004}. Some immigrants do this by carrying tangible objects like family photographs and traditional artifacts from home \cite{sabie_memory_2020}, while others engage in community activities with fellow immigrants from home \cite{bhugra_migration_2004}. 

   Anderson’s notion of “long-distance nationalism” captures how immigrants maintain political, cultural, and emotional attachments to their homeland through modern communication technologies, which facilitate information exchange, media consumption, and remote participation in homeland affairs \cite{anderson_spectre_1998}. A closely related term, transnationalism, is defined by Shklovski as the dynamic, ongoing processes of connection and identity formation that transcend national borders \cite{shklovski_introduction_2014}. Among such transnational practices, remittance \cite{noauthor_remittance_2025} flows have received substantial scholarly attention, including recent studies of Iranian immigrants in Canada \cite{yousefi_remittance_2025}; however, nonprofit and volunteer-based contributions remain comparatively understudied.
    
    Homeland-oriented nonprofit activity, such as sending monetary and intellectual help to underserved communities, provides immigrants with a sense of purpose and belonging in their new country; and supports the reconstruction of shaken cultural identities, empowerment, and improving psychological well-being \cite{martinez-damia_community_2023, sveen_immigrants_2023, ouacha_diasporic_2024, baig_re_2016}. Such engagement often stems not only from identity repair but also from a sense of moral responsibility, emotional ties, and aspirations for the development and well-being of the homeland \cite{baker_diaspora_2013, flanigan_crowdfunding_2017, paula_d_johnson_diaspora_2007}.

     Despite the major impact of nonprofit activities on immigrants' lives, broader political and institutional conditions may shape whether these contributions are recognized and supported \cite{ouacha_diasporic_2024, baig_re_2016}. They are constrained by substantial structural and organizational barriers, including bureaucratic obstacles such as customs procedures, limited information platforms, and unreliable local data \cite{desouza_transnational_2023, newland_diaspora_2010, brinkerhoff_diaspora_2008}. At the organizational level, immigrant-based nonprofits additionally contend with chronic funding shortages, volunteer instability, coordination difficulties, and limited institutional legitimacy \cite{martinez-damia_mediating_2024}.

    Previous research has shown that while information technologies respond to immigrants' settlement needs, they minimally address their cultural, emotional, and financial needs \cite{hsiao_technology_2018}. Furthermore, despite the centrality of digital communication to these transnational nonprofit practices \cite{desouza_transnational_2023}, prior scholarship rarely examines how immigrant-led nonprofit work interacts with technological infrastructures. This gap is increasingly consequential as communities come to expect technology-supported interactions with nonprofit organizations. We next examine research on nonprofit technologies in HCI to contextualize how digital systems support, constrain, or reshape nonprofit work.

    \subsection{Cooperative Work and Alternative Organizing at the Margins}
    \label{nonprofit sociology}

    Immigrant-led nonprofit groups rarely resemble the stable, bureaucratic organizations assumed in dominant nonprofit technology design. Instead, they operate within constrained socio-technical environments shaped by volunteer labor, informal governance, and cross-border political and logistical pressures. Scholarship on nonprofits working at the margins offers insight into how these structural conditions shape everyday organizational practices.
    
    These groups navigate unique constraints related to their reliance on volunteer labor \cite{godefroid_identifying_2023}, transnational structures \cite{shklovski_introduction_2014, williams_multisited_2014}, limited technical and financial resources, and cultural, linguistic, and access barriers \cite{hsiao_technology_2018}. These groups depend heavily on social capital to compensate for systemic limitations \cite{king_social_2004}, and maintaining their infrastructures, encompassing organizational routines, communication systems, and social networks, requires constant work \cite{sackitey_virtual_2025, merkel_managing_2007, karasti_infrastructuring_2014}. This labor includes substantial articulation work \cite{suchman_supporting_1995,schmidt_taking_1992} and ongoing legitimacy work to gain credibility with donors, governments, and community members \cite{tanaka_legitimacy_2016}.

    These observations connect immigrant nonprofit practice to foundational CSCW concerns. Prior workplace articulation work, defined as the continuous alignment of interdependent tasks, actors, and resources, as the central problem cooperative systems must support~\cite{schmidt_taking_1992}, and shows how groups build coordination mechanisms, from protocols to shared artifacts, to reduce its cost~\cite{schmidt_coordination_1996}. Cooperative work further depends on mutual awareness of others' activities~\cite{dourish_awareness_1992}and on common information spaces in which distributed actors maintain shared records and interpretations~\cite{bannon_constructing_1997,schmidt_taking_1992}; sustaining these across distance is notoriously difficult, as remote collaborators lack the common ground that collocation provides~\cite{olson_distance_2000}. Much of this work, finally, is structurally invisible~\cite{star_layers_1999}, unrecognized and therefore unsupported by formal systems. Volunteer-run immigrant nonprofits inherit all of these problems in acute form~\cite{yousefi_remittance_2025,rohanifar_money_2021}: their coordination mechanisms are improvised atop consumer messaging apps, their common information spaces are homebrew and fragile~\cite{voida_shapeshifters_2011,voida_homebrew_2011}, and the actors who perform articulation work are unpaid and transient.

    In response to these pressures, many grassroots and immigrant-based nonprofits adopt informal, improvised, and flexible solutions as mechanisms of resilience \cite{yousefi_remittance_2025, nguyen_it_2025}. Such informality enables adaptability and responsiveness \cite{stone_planning_1996} but introduces vulnerabilities related to legitimacy, visibility, and scalability \cite{chandra_informality_2017}. As a result, immigrant-run nonprofits experience intersectional marginalization \cite{wong-villacres_designing_2018}, constrained by both their immigrant status and the nonprofit nature of their work, with limited access to institutional, financial, or civic support structures.
    
    These dynamics make informal, adaptive organizing practices not peripheral but foundational. The scholarship on alternative forms of organizing deepens this understanding by illustrating how structurally excluded communities, such as undocumented migrants, diasporic groups, or politically sensitive nonprofits, develop organizational forms that deviate from dominant institutional templates when formal pathways are inaccessible, risky, or misaligned with community needs. These insights are particularly relevant for Iranian immigrant nonprofits, which must navigate sanctions, geopolitical risks, and credibility challenges while serving dispersed, multilingual communities.
    
    Models of alternative organizing documented among groups such as low-income urban residents and migrant women’s cooperatives emphasize community-driven values, collective labor, flexible governance, and improvised structures \cite{cheney_alternative_2017, vijay_another_2025, imas_harare_2012, zapata_campos_expansion_2024, avalos_mutual_2019, choudhary_locating_2024}. Rather than being disorganized, these groups continuously manage tensions between visibility, credibility, autonomy, and safety. For example, organizations may strategically limit their public presence to protect members while cultivating sufficient transparency to maintain trust among supporters \cite{jensen_alternative_2015, kangas-muller_doing_2024}.
    
    Digital technologies increasingly play a central role in such alternative organizing. Studies of mutual aid networks demonstrate how communities construct hybrid digital–physical infrastructures to coordinate care, share information, manage risk, and sustain collective identity when formal institutions fall short \cite{wilson_mutual_2022, hultquist_digital_2022, yantseva_immigrant-critical_2023}.
    
    Together, prior research shows that alternative organizing constitutes a common, adaptive mode of collective action among marginalized communities, and that digital tools critically shape how these groups coordinate, remain safe, and maintain legitimacy.


\subsection{Technology, AI, and HCI in Nonprofit Practices}
\label{nonprofit technology}
Nonprofit organizations rely on a wide spectrum of digital tools, including websites, social media, third-party platforms, crowdfunding systems, donor relationship management tools, and data management infrastructure, to support communication, donor engagement, fundraising, evaluation, advocacy, and capacity building. HCI and adjacent fields have examined these technological practices, highlighting both their potential and their limitations \cite{goecks_charitable_2008}. Information systems research often foregrounds opportunities for AI-enabled interventions, such as personalized fundraising \cite{rishabh_singh_empowering_2022}, persuasive donation assistants \cite{song_would_2024}, or algorithmic classification of immigrant-serving nonprofits \cite{ren_new_2022}. However, scholars emphasize the need for such innovations to be attentive to the lived realities and human dimensions of nonprofit work~\cite{kanter_smart_2022} as organizational adoption depends not only on functionality but also on human and structural factors, including chronic under-resourcing, limited technical capacity, organizational culture, and legal constraints \cite{godefroid_identifying_2023}. Similarly, recent advances in large language models (LLMs) and conversational AI systems have introduced new possibilities for mediating interactions between such organizations and their stakeholders: automating communication, generating reports, and supporting decision-making. However, prior work has also highlighted myriad ways they inherit and exhibit biases from underlying data (e.g., under-representation and stereotypes)~\cite{gallegos_bias_2024, shrawgi_uncovering_2024}, reproduce misinformation~\cite{garry_large_2024, pan_risk_2023}, and fail to account for the situated, relational, and trust-sensitive nature of nonprofit work~\cite{bender_dangers_2021}, concerns that are often of utmost importance in this context.

A parallel body of work examines AI not as a tool but as a collaborator. Research on machines as teammates maps how AI systems may take on coordination, evaluation, and decision-support roles within human teams, along with the risks such roles introduce~\cite{seeber_machines_2020}; empirical CSCW studies show that practitioners integrate AI into existing collaborative workflows selectively, treating it as a junior partner whose outputs require verification and whose role must be negotiated~\cite{wang_human-ai_2019,cai_hello_2019}. Recent work extends this line to LLM-based agents that participate directly in group activities such as collaborative ideation~\cite{he_ai_2024}. This literature, however, centers well-resourced professional teams like data scientists, clinicians, and knowledge workers. How AI participates in the cooperative work of volunteer-run, resource-poor, and politically vulnerable collectives, where trust is the binding resource and human relationships carry the organization, remains largely unexamined, and it is the question our design and evaluation of AMINA takes up.

Beyond core organizational technologies, HCI research has addressed broader facets of nonprofit and social-sector work: designing with dignity for homeless populations \cite{le_dantec_designs_2008}, supporting the formation of civic communities \cite{le_dantec_infrastructuring_2013}, enabling care work in everyday residual moments \cite{harmon_supporting_2017}, supporting activists whose work is delegitimized by authorities \cite{mariam_asad_illegitimate_2015}, and examining informal interactions within nonprofit networks \cite{stoll_informal_2010}. Across this work, HCI scholars critically foreground the structural challenges that shape technology use in nonprofits. Early studies highlighted uneven technical capacity and the ongoing negotiations, driven by fluid work requirements, limited resources, and usability constraints, required to sustain technology under conditions of organizational precarity \cite{merkel_managing_2007, voida_shapeshifters_2011, voida_homebrew_2011}. Despite these barriers, as data-driven work becomes increasingly central to organizational life, nonprofits deploy data and technology in ways that remain deeply entangled with everyday challenges.
    
A recurring theme in the literature is the gap between real-world social work practices and the tools designed for nonprofit use ,many of which rely on “design fictions” rather than actual user needs, leading to underperformance or added administrative burdens \cite{harmon_design_2017}. In response, nonprofit workers frequently construct improvised “homebrew databases,” shape-shifting socio-technical assemblages sustained through peer learning and iterative improvisation rather than formal IT infrastructures \cite{voida_homebrew_2011, voida_shapeshifters_2011}. Related work on data-advocacy tools shows how community and advocacy organizations appropriate data to contest policy and shape public discourse while navigating asymmetries in technical capacity and infrastructural support \cite{boone_designing_2025}. Additional studies examine how technology reshapes grassroots organizational values \cite{nguyen_it_2025} and how mutual aid groups assemble socio-technical infrastructures during crises \cite{soden_dilemmas_2021}, confronting dilemmas around inclusion, urgency, formality, sustainability, privacy, transparency, solidarity, and co-production \cite{soden_dilemmas_2021, nguyen_it_2025}.

Data-driven expectations can also impose operational strain on nonprofits, diverting attention from core missions toward producing metrics for external stakeholders; an effect described as “mission drift” \cite{bopp_disempowered_2017}. Scholars have proposed data-handling frameworks tailored to nonprofit realities to mitigate these unintended burdens \cite{gondimalla_aligning_2024}. Darian et al. further conceptualize data in four roles, namely, legitimizer, activator, amplifier, and incubator, highlighting both its potential for public good and the challenges of working within marginalized data economies, while calling for data feminist approaches to address power asymmetries \cite{darian_enacting_2023}. Relatedly, Erete et al. show that nonprofits use data to craft narratives for funders, but emphasize that data alone is insufficient; tools should help translate data into meaningful stories \cite{erete_storytelling_2016}.

HCI scholarship also includes several meta-analyses of nonprofit and social-sector technologies. Sackitey et al. document how digital platforms simultaneously enable rapid coordination and generate new burdens for mutual aid organizers, including burnout, conflict, exclusion, and competing priorities \cite{sackitey_virtual_2025}. Bopp et al.’s systematic review identifies systemic biases shaping research trends and introduces the concept of “analytical charisma” \cite{bopp_voices_2020}. Chordia et al. expand this perspective through a comprehensive review of social-justice work in HCI, mapping strands such as racial, disability, environmental, economic, and design justice, and urging deeper engagement with justice-oriented approaches \cite{chordia_social_2024}. Complementing these academic analyses, Hardy et al. call for asset-based approaches to community-based computing, shifting away from deficit narratives \cite{hardy_turn_2024}. Other work examines nonprofit technology practices on the ground: Erete et al.~\cite{erete_towards_2025} advocate for equitable community–industry collaborations that confront power asymmetries, and recent studies of “advocacy technologists” highlight the precarious, often informal labor involved in translating between policy, advocacy, and technology within nonprofits \cite{chambers_beyond_2025}.
    
Collectively, this body of HCI and CSCW research calls for designing inclusive systems that explicitly attend to the socio-political infrastructures and structural barriers shaping nonprofit work, rather than uncritically applying state-of-the-art technologies. While prior studies highlight both the promise of technological support and the persistent mismatch between nonprofit realities and available tools, they give limited attention to organizations operating at the intersection of migration-related precarity \cite{sabie_decade_2022}, geopolitical constraints, and informal infrastructures, which is the focus of this paper.

\subsection{Design Justice}
To address these complexities, design frameworks that foreground community power, structural inequity, and lived expertise are needed. Design justice provides a foundation for assessing whether technologies, especially AI-driven ones, support or undermine the organizational realities of marginalized nonprofits.

While prior scholarship has examined immigrants’ nonprofit activities through sociological and psychological lenses (Sections \ref{immigrant identity}, \ref{nonprofit sociology}) and HCI work has investigated technology use in nonprofit settings (Section \ref{nonprofit technology}), the intersection of these domains remains underexplored, particularly the group-organized practices of immigrants operating in politically sensitive contexts such as Iranian nonprofits in the Global North.

Guided by design justice, we treat design as a site where benefits and burdens are distributed and often reproduce broader matrices of domination, including racialized inequality, legal vulnerability, geopolitical exclusion, and knowledge hierarchies \cite{costanza-chock_design_2020}. Rather than presuming technological neutrality or universalist models of “nonprofit,” this perspective centers lived expertise and highlights how geopolitical tensions and migration-related precarity shape nonprofit realities.

For IINGs, this requires careful attention to how technologies mediate visibility, legitimacy, and access to resources \cite{tanaka_legitimacy_2016, hsiao_technology_2018}. Design justice asks whose narratives and organizational practices are obscured when platforms or policies ignore marginal nonprofits. This question extends CSCW's long-standing attention to invisible work: rendering informal cooperative labor legible to systems, without exposing its performers to new risks, is both a justice problem and a classic design problem, as visibility can bring recognition, but it can also bring surveillance~\cite{star_layers_1999}.

In this study, we operationalize this lens by: (1) Understanding the socio-technical realities of IINGs through in-depth interviews; (2) engaging stakeholders in co-designing the goals and behaviors of an AI assistant; and (3) evaluating the resulting system not only for efficiency-related outcomes but also for its inclusiveness, accountability, and alignment with community control \cite{laura_ramirez_galleguillos_how_2020, seguin_co-designing_2022, vines_configuring_2013}.

Building on these perspectives, participatory and co-design approaches have been widely adopted in HCI and group work scholarship as methodological instantiations of design justice, enabling communities to actively shape technologies that affect their lives rather than serving as passive and docile subjects of design and technological interventions~\cite{vines_configuring_2013, seguin_co-designing_2022}. In this paper, we employ co-design not only as a method for eliciting requirements but as a means of redistributing design authority to immigrant nonprofit practitioners and organizers, whose situated knowledge is crucial for shaping AI systems that align with their operational and political realities. Following traditions of research-through-design in HCI, we further operationalize these principles by prototyping and evaluating an AI assistant that embodies the identified design goals. This approach enables us not only to conceptualize but also to empirically examine whether and how such systems may support or challenge existing nonprofit practices.

\section{Methods Overview}
\label{methods overview}
This section provides an overview of the methods in this multi-phase study, with methodological details of each phase presented in their respective sections. The study employed a three-part mixed-methods approach: (1) semi-structured interviews with nonprofit leaders and active members of IINGs, (2) a co-design session, and (3) evaluation and feedback sessions on the prototyped AI assistant, AMINA. The IINGs were located in Canada, Europe, the UK, and the USA. While Iranian nonprofit activity exists globally, we focused on these regions due to the authors’ long-standing ties, research access in these regions. These countries also represent key sites of AI deployment and established immigrant infrastructures, making them relevant to our research question. We distinguish the studied nonprofits by their level of formality. Formal groups are government-registered and can issue tax-deductible receipts. Semi-formal groups are registered with institutions such as universities or community centers. Informal groups, on the other hand, lack legal registration but maintain some organizational structure.

All researchers in this study are immigrants in North America. Also, their disciplinary background is in HCI and social computing, with a focus on designing technology and infrastructure for marginalized communities. They are also actively involved in charitable activities, with the primary researcher bringing over a decade of experience, including roles such as group leader, HR manager, public outreach coordinator, technology manager, and content creator. Her extensive field experience, spanning work with beneficiaries and collaboration with over 20 charity groups, informed the interview design and analysis and enabled the integration of contextual insights into this article, including explanations about specific tools and platforms used in the nonprofit sector.

Participants across all phases of this study were recruited via study flyers posted on the first author’s Instagram and Telegram channels. Recruitments followed a snowball sampling approach \cite{biernacki_snowball_1981}, where initial participants referred additional members from their networks. This recruitment strategy was particularly suitable for reaching participants involved in informal and transnational charitable activities, which are often difficult to identify through institutional recruitment channels. This study received ethics approval from the authors’ institutional review board, and participation was voluntary and unpaid.

Across all three phases, we engaged IING leaders, volunteers, and frequent donors primarily based in the Global North, along with several Phase 2 participants residing in Iran who collaborate with cross-border charity groups. Phase 1 consisted of 27 semi-structured interviews examining how participants and their organizations interact with technology and AI in the context of transnational charitable work. Insights from Phase 1 informed the design of Phase 2, a 120-minute online co-design session where participants collaboratively generated expectations, personas, and design features for a nonprofit AI assistant using a shared Miro board. Based on Phase 2 outputs, we prototyped the AI assistant AMINA. In Phase 3, we conducted one-on-one evaluation and feedback sessions with 7 nonprofit practitioners, in which participants interacted with AMINA via a Figma-based interface, completed donor and volunteer-oriented task flows using a think-aloud protocol, and reflected on their impressions, experience, usability, and broader design implications of the prototype. 

All sessions were conducted in Farsi over Zoom. Across phases, transcripts, notes, and co-created materials were analyzed through inductive thematic analysis using a common procedure detailed in~\ref{phase1_analysis}: independent initial coding by two authors, consensus meetings supported by a maintained codebook, and affinity-diagram clustering into themes. Themes were consolidated across phases in team meetings, with Phase 1 themes serving as sensitizing concepts for later phases; disagreements throughout were resolved through discussion to consensus~\cite{mcdonald_reliability_2019}. In the following sections, we detail the methodology and findings for each phase.

\section{Phase 1: Understanding Charitable Activities in the Iranian Immigrant Population}

    The first phase of our study consisted of semi-structured interviews with executive members and volunteers of IINGs. The goal of this phase was to examine (RQ1): "How do Iranian immigrant nonprofit groups coordinate charitable work across formal and informal arrangements, and how do existing digital infrastructures play a role in their interaction?" In this section, we present the details of methods, findings, and a discussion of the key takeaways from the findings of phase 1. 

    \subsection{Methods}
    
    \subsubsection{\textbf{Participant Recruitment}}
     In the first phase, semi-structured interviews, we recruited 27 leaders or active members from 17 distinct IINGs in the diaspora located in Canada, Europe, the UK, and the USA. Tables \ref{tab:demog1} and \ref{tab:demog2} present the demographic and background information of participants and their nonprofit group of activity. Please note that most organizations provide service in more than one area of support and more than one type of technology.

        \begin{table*}[!htbp]
    \centering
    \caption{Distribution of Age, Gender, and Occupational Background Categories Among Interview Participants}
    \label{tab:demog1}
    \small
    \begin{tabular}{|c|c|c|}
    \hline
    \multicolumn{3}{|c|}{Number of Participants: 27 (Female: 12, Male: 15)} \\
    \hline
    Age Range (in Years) & Gender & Occupation Background of Participants \\
    \hline
    20--30: 5 & Female: 12 & Engineering: 14 \\
    30--40: 11 & Male: 15 & Management \& Business Administration: 5 \\
    40--50: 6 &  & Health Sciences: 4 \\
    50--60: 3 &  & Psychology \& Education: 2 \\
    60--70: 2 &  & Art \& Design: 1 \\
              &  & Finance \& Accounting: 1 \\ 
    \hline
    \end{tabular}
\end{table*}


\begin{table*}[!htbp]
    \centering
    \caption{\centering Distribution of Participants' Nonprofit Groups Across Nonprofit Types, Service Areas, and Technology Use \footnotesize{(* Please note that some groups are focused on more than one area of aid and support)}}
    \label{tab:demog2}
    \small
    \begin{tabular}{|>{\centering\arraybackslash}p{1.6cm}|>{\centering\arraybackslash}p{1.8cm}|>{\centering\arraybackslash}p{5.3cm}|>{\centering\arraybackslash}p{4.9cm}|}
    \hline
    \multicolumn{4}{|c|}{Number of Distinct Nonprofit Groups: 17} \\
    \hline
    Formality Level & Size of Nonprofit Group & Areas of Aid and Support  & Technologies Currently Used by Nonprofit Group (Example) \\
    \hline
    Formal: 5 & 1--10: 2 & Education \& Literacy Support: 7 & Social Media (Facebook, Instagram, ...): 12  \\

    Semi-formal: 6 & 10--20: 4 & Health Assistance \& Medical Infrastructure: 8 & Donor-Relationship Managers (Bloomerang, Bonterra): 3\\
    Informal: 6 & 20--50: 8 & Nutrition \& Food Security: 7 & Messaging Apps (Telegram, WhatsApp): 17 \\
                 & 100--200: 2 & Refugee \& Newcomer Settlement Support: 3 & Website (Wix, WordPress): 13 \\
                 & 300--400: 1 & Community Development: 5 & Financial Technology (PayPal, Donorbox, Email Money Transfer): 17 \\
                 &             & Empowerment \& Career Development: 4 & Event Management Platforms (Eventbrite, Zeffy):6 \\
                 &             & Crisis Response \& Humanitarian Relief: 4 & Accounting Software (QuickBooks, Custom-made Software): 4  \\
                 &             & Drinking Water Access: 1 & Collaborative Information Management Clouds (Google Drive): 17 \\
                  &             &  & Content Creation Platforms (Canva, Adobe): 16 \\

    \hline
    \end{tabular}
\end{table*}

     \subsubsection{\textbf{Data Collection}} The data collection was conducted through semi-structured interviews, where the researchers' backgrounds informed the development of preliminary questions. Participants were asked about their group's mission and management structure, fundraising methods and payment gateways, social media interaction, communication with the audience and beneficiaries, collaboration with other nonprofit groups, volunteer coordination, project management, long-distance nonprofit work, the challenges associated with each of these aspects, and the role of technology and Artificial Intelligence in either facilitating or exacerbating them.  While the interviews began with a set of prepared questions, the discussions were guided by the flow of the conversation, allowing for flexibility and depth in exploring relevant topics. The semi-structured interviews uncovered multi-layered, context-specific realities of charitable work, while allowing flexibility to explore unexpected aspects. In total, we conducted 27 sessions of interviews (38.5 hours with an average length of 86 minutes) and 351 pages of transcription. 
     
     \subsubsection{\textbf{Data Analysis}}
     \label{phase1_analysis}
     Data analysis involved a rigorous review of interview transcripts. An iterative process was employed, during which each transcript was carefully reviewed to isolate and emphasize segments relevant to our research questions while eliminating irrelevant portions. Frequent virtual team meetings were held to discuss and verify the excluded segments, ensuring no significant excerpts were overlooked. Following this initial review, the anonymized data were analyzed using open coding \cite{strauss_basics_1990} and thematic analysis \cite{boyatzis_transforming_1998}. This approach was deliberately inductive, avoiding predetermined themes. Two authors independently open-coded a shared subset of 12 transcripts, then met bi-weekly to compare codes and maintain a shared codebook of exemplar excerpts. Disagreements, most often over the boundary between adjacent codes, were resolved through discussion to consensus, returning to the transcript context and drawing on the first author's field experience where interpretation was culture- or sector-specific. The remaining transcripts were coded by the first author, with ambiguous excerpts flagged for group discussion. Codes were clustered into themes via affinity diagramming in full-team sessions, retaining themes supported across multiple participants~\cite{mcdonald_reliability_2019}.

    \subsection{Findings of the Interview Study}
        \label{interview findings}
        In this section, we draw on the analysis of interview data to present the findings, which demonstrate the broad optimism of participants regarding the transformative potential of AI for enhancing efficiency in nonprofit work, as well as challenges that temper this optimism, and caution against the uncritical deployment of data-driven systems on top of already non-inclusive and misinformed digital infrastructures. 
        Despite participants' consistent recognition of AI’s promise, especially given the resource and labor constraints common in these nonprofit groups, the analysis of interview data reveals two interconnected layers of challenge: First, the fractured socio-technical infrastructures underpinning their work, characterized by systemic exclusion, capacity limitations, and persistent misinformation; and second, the broader structural marginalization of IINGs within mainstream technological ecosystems, along with the circulation of misinformation about these groups across online environments. 

        Before elaborating on these two layers in the next subsections, we first characterize the general workflow of IINGs, and the cooperative work these challenges disrupt: the collaborative arrangements through which IINGs accomplish charitable action, which form the ground against which exclusion, capacity gaps, and misinformation acquire their significance. Charitable action in IINGs is a collective accomplishment sustained by a distributed and largely voluntary division of labor. A typical campaign links (i) executive members in the diaspora who set direction and hold financial accountability of charitable projects (ii) volunteers who run outreach, event logistics, and donor communication, who might live in a different geographical location in diaspora (iii) trusted partner social workers and organizations inside Iran who identify beneficiaries, and document the delivery of financial aid and intellectual support, (iv) donors, across the globe, who expect ongoing evidence that their contributions arrive as intended, and (v) similar groups and organizations in diaspora with shared purpose who sometimes engage in joint projects. Because these actors are separated by distance, time zones, language, and legal regime, the work exhibits the classic difficulties of distributed collaboration~\cite{olson_distance_2000}: participants of this transnational collaboration network must establish common ground and mutual awareness with few shared artifacts and no co-located encounters. Holding this arrangement together demands substantial articulation work ~\cite{suchman_supporting_1995,schmidt_taking_1992}, meaning the meshing of tasks, actors, and resources that keeps cooperative work on track, much of it invisible~\cite{star_layers_1999}. Much of this articulation work is performed through messaging apps, especially Telegram, and shared Google drives, and other general-purpose IT services. Under the conditions of sanctions and surveillance, these systems serve as improvised coordination mechanisms~\cite{schmidt_coordination_1996,bannon_constructing_1997}: they are where people and project management are achieved, and where they break down.

        With this workflow in mind, the following sections provide a detailed elaboration of the two layers mentioned above.

        \subsubsection{\textbf{Fractured Socio-Technical Infrastructures in Charitable Practices of the Iranian Diaspora}}
        \label{social infrastructure}
         
        \begin{enumerate}
            \item \textbf{Legitimacy Barriers and Structural Exclusion from Official Registries}
            \label{Exclusion, and the Hard Path of Obtaining Legitimacy}
    
            A substantial portion of IINGs operate informally outside structures recognized by government. Formalizing their status is particularly difficult: complex international politics surrounding Iran introduce scrutiny and delays into otherwise standard legal procedures, and groups supporting communities in Iran face additional barriers due to sanctions and geopolitical tensions. These challenges, combined with limited volunteer capacity, scarce labor, and financial resources, make legal registration burdensome. As a result, many groups remain excluded from nonprofit-specific infrastructures such as directories, fundraising and banking platforms, and corporate donation-matching programs, all of which typically require formal status. To remain functional, they rely on workarounds such as accepting donations through cash or personal accounts, or partnering with intermediaries to gain visibility, issue tax receipts, and manage international transfers. While these strategies enable continued operation, they constrain scale and sustainability.
    
            Participant 6 is a lifelong activist who collaborates with a nonprofit in Iran that supports vulnerable children from abusive families. After immigrating, she ran fundraising campaigns for the organization and sought government registration to access digital financial platforms, social media fundraising, and corporate fund-matching programs. As an illustration of these challenges, she recounted:  
    
            \begin{quote}
            \textit{"We talked to a lawyer about registering, but she told us that once ‘Iran’ appeared on the file, requirements became stricter and more complicated. The requirements were beyond our time and funding limits, so we decided not to move forward with group registration."} (P6)
            \end{quote}
        
            Similarly, P1, who manages a university-based nonprofit, noted that their PayPal account was blocked after donors mentioned “Iran” in comments. To prevent further disruptions, the group began a formal registration process five years ago, which at the time of the interview remained incomplete. These examples illustrate how structural barriers to legitimacy uniquely challenge groups from attaining government recognition and continuing their activities informally. 
            
            \item \textbf{Capacity Gaps in Technology Adoption}  
                Our findings reveal capacity gaps in technology adoption, stemming from limited resources, language and cultural barriers, digital literacy gaps, and trust and safety concerns. These mirror barriers, identified separately in studies of immigrants \cite{hsiao_technology_2018} and nonprofits \cite{godefroid_identifying_2023}. At the intersection of these identities, however, IINGs also face high volunteer turnover, which disrupts continuity and hinders technology adoption. As P3 explained:

                \begin{quote}
                    \textit{"Many volunteers are newcomers struggling with English and unfamiliar with available resources. Once they settle, they often move on to other jobs, taking their knowledge with them. For example, a volunteer once suggested the Amazon Smile [donation matching] program, but left before we could follow up."} (P3)
                \end{quote}
                    
                These dynamics leave many groups unable to adapt to evolving nonprofit-specific digital platforms and support structures.
            
            \item \textbf{Misinformation}  
            \label{misinformation}
            A recurrent theme in participants’ accounts concerns the risk of becoming the target of false accusations regarding political affiliations or the recipients of their financial aid. Many Iranians in the Global North are critics of the Iranian regime; consequently, suspicion toward any group perceived to have financial ties to Iran is high. Participants reported that, irrespective of factual accuracy, accusations of collaboration with state-affiliated entities were often circulated through social media and messaging platforms, creating reputational harm and internal disagreements within their groups. As P9 recounted her experience:  
            
            \begin{quote}
            \textit{"During Mahsa Amini movement \cite{noauthor_mahsa_2025}, all Iranians were angry. Then I saw in a Telegram post saying our charity has sent \$40{,}000 to [a religious institute]. We were not even remotely close to that Institute. These incidents put heavy mental pressure on our teams."} (P9)
            \end{quote}
            
            Participants further explained that accusations were more likely to escalate when group members displayed visible signs of religiosity. Some organizations responded by limiting the public visibility of such members to avoid being perceived as politically aligned with the Iranian State:  
            
            \begin{quote}
            \textit{"We are a diverse group, but usually ask our friends with headscarves, even our president, not to show up at events or social media; otherwise, it would trigger so many questions."} (P17)
            \end{quote}
            
            Other participants reported taking proactive precautionary measures to preempt the misinformation. Such measures included explicitly stating on their websites that they are non-religious and unconnected to any political entity. They also emphasized frequent, detailed financial reports and photo documentation demonstrating that all donations were merely allocated to aid recipients.  
            
            \begin{quote}
            \textit{"We expected that we might be subject to attacks regarding connections to Iran's government. So we gave frequent reports and were vocal about not being connected to any party. We keep activities low-key to avoid complications, which may also have limited our growth."} (P21)
            \end{quote}
            
            These accounts demonstrate how political stigmatization, fueled by misinformation networks, shapes organizational strategies. Groups frequently resort to strategic invisibility \cite{star_layers_1999} and rigorous transparency protocols to safeguard the continuity of their mission.  
            
        \end{enumerate}

        \subsubsection{\textbf{Interactions with Technology and AI in the Shadow of Socio-Technical Infrastructures}}
        \label{findings-infrastructure-shadow}
            While Section~\ref{social infrastructure} focused on the socio-technical infrastructures of IINGs, this section presents participants’ engagement with computing technology in daily operations, alongside their perspectives on emerging AI in this context.  
            
            \begin{enumerate}

            \item \textbf{The Need for Information Technology Support}
            \label{findings-tech-need}
            
            All nonprofit practitioners in this study emphasized the significant role of information technology in their work and expressed a clear desire for tools that better support their organizational tasks, while remaining mindful of the volunteer-driven and resource-constrained realities of their context. Participant 13 described the value of digital systems as follows: 
                \begin{quote}
                    \textit{"Technology has played a tremendous role in most aspects of our work. For example, a truck driver from the villages of Lorestan [in Western Iran] approached us and requested assistance. Would it be possible without social media? Recently, I used an LLM to draft a document in two hours. In the past, I would have spent two months going back and forth with lawyers to produce the same document. None of this would have been possible without AI...There is enormous potential for customizing AI for nonprofit work."} (P13)
                \end{quote}

            Participants' expectations of digital technologies were correspondingly collaborative rather than merely individual. Beyond personal productivity, they wanted tools that would preserve organizational memory across volunteer turnover, keep donors and volunteers mutually aware of a campaign's state, coordinate with partner groups and intermediaries, and reduce the articulation burden of transparency work, producing, in effect, the coordination support that registered nonprofits obtain from dedicated platforms. For example, P21 described relying on a \textit{homebrew} digital system, which preserves organizational memory via documentation of data and communications, and manages donor-relationship~\cite{voida_homebrew_2011}, which is a critical, yet fragile, part of their operations:

                \begin{quote}
                    \textit{"We asked a volunteer who was a software engineer to build a system that automates accounting, issues receipts, manages donor profiles on the website, etc. We depend heavily on this system, but it breaks down fairly often, and the volunteer is too busy to maintain it as regularly as we need."} ({P21})
                \end{quote}

                She further expressed hope that new AI tools might help volunteers handle a greater volume of tasks. These expectations frame the challenges reported above: legitimacy barriers, capacity gaps, and misinformation are consequential precisely because they drain labor and resources, corrode the trust on which distributed voluntary collaboration depends, and damage the cooperative fabric.

                 Taken together, these accounts highlight the need for context-aware, sustainable information services that can support nonprofits operating at the margins.

            \item \textbf{Second-order Exclusion from Digital Services}  
            \label{findings-2nd-order-exclusion}
            Despite the critical role of technology, our findings reveal that a substantial number of IINGs, particularly informal and semi-formal ones, are systematically excluded from the benefits offered by available nonprofit-specific digital services due to the difficult path toward legitimacy. Many corporate platforms offer specialized features or financial benefits for nonprofits, yet these are contingent upon government-validated registration. For example, nonprofit directories such as Charity Navigator \cite{noauthor_charity_2025} and GuideStar \cite{noauthor_guidestar_2025} only recognize officially registered organizations. Similarly, fundraising services on Facebook and Instagram \cite{noauthor_meta_2025} are restricted to formally registered nonprofits. Other examples include fee waivers, grants, and corporate donation matching programs \cite{noauthor_benevity_2025}, such as Microsoft 365 Nonprofit~\cite{noauthor_solutions_2025}, PayPal Giving Fund~\cite{noauthor_paypal_2025}, and Stripe fee waivers~\cite{noauthor_stripe_2025}, all of which are specific to registered nonprofits. Even platforms explicitly designed for nonprofit sectors, such as Zeffy \cite{noauthor_zeffy_2025}, do not recognize informal nonprofit practices. P10, a member of a semi-formal charitable group, described the impact of these exclusions:  
            
            \begin{quote}
            \textit{"We use Donorbox~\cite{noauthor_donorbox_2025} and Stripe~\cite{noauthor_stripe_2025} for setting up our online donation. They take processing fees. So, we encourage donors to send direct money transfers to our bank accounts because we want to be loyal to the donors' money. For event ticketing, we use free services like Google Forms, ask attendees to send the fees through direct transfer, and then a volunteer manually matches the transactions and sign-ups. There's always a discussion about this in the team."} (P10)
            \end{quote}
            
            As also noted in Section~\ref{social infrastructure}, capacity gaps, too, further constrain the adoption of digital services, which results in diminished digital footprints and interaction with AI tools. P17 explains about their organization which is mostly managed by senior members:
            
            \begin{quote}
            \textit{"We use [mobile] applications in our group, but the use is limited. I even sometimes use ChatGPT, but not all members are comfortable. When I mention a software, they say wait, I'll ask my son to install it for me."} (P17)
            \end{quote}
            
            Similarly, P23 described the challenges of working with digital tools amid limited labor resources and expressed optimism that AI might eventually take on some of these tasks. At the same time, she acknowledged the organization’s limited online presence:
            
            \begin{quote}
            \textit{"We have provided drinking water systems in tens of remote villages and sent nutrition and drugs to thousands of kids who were dying from malnutrition. But we don't usually get the time to put the reports on social media. Maybe AI does a better job of reporting our impacts. But how does AI find us? I wonder what we can do to help AI [find us]?"} (P23)
            \end{quote}
            
            These examples reveal that many impactful charitable endeavors occur outside the digital sphere, leaving little to no online trace. This absence is not incidental but rather a direct consequence of the fractured design of digital tools and services, which assume formal registration and ignore capacity gaps. The result is a form of practical technological exclusion: under-resourced and unregistered nonprofits remain digitally invisible, thereby reducing their discoverability in emerging data-driven AI systems. This cascade effect will be further elaborated in section \ref{interview discussion}.  
            
            \item \textbf{Online Misinformation and Penetration into AI-powered tools}  
            \label{findings-2nd-order-misinfo}
            The social infrastructures mentioned in Section~\ref{social infrastructure}-\ref{misinformation} unpacked the misinformation and false accusations targeting IINGs. Findings also show the implications of this reality on their digital profiles, raising concerns about AI tools potentially amplifying such distortions. P18, a long-time nonprofit volunteer, compared experiences of misinformation in online and offline contexts:  
            
            \begin{quote}
            \textit{"The inherent characteristics of social media allow some users to hide behind fake usernames and propagate false information. They accuse us of money laundering, which is extremely heavy on us. We do see hate speech in person, too, but the degree of accusations online is much further from reality."} (P18)
            \end{quote}
            
            P19, a computer engineer managing a semi-formal group, expressed similar concerns:  
            
            \begin{quote}
            \textit{"The online data about charities is mixed with incorrect information. This type of conversation often gets out when a controversial incident happens that heats up the news. We are not connected to anyone. I'm worried about how AI is going to deliver this information."} (P19)
            \end{quote}
            
            These accounts underscore the interplay between misinformation and digital infrastructures. Social networks not only trigger and propagate false narratives but also shape the datasets that underpin AI systems. 
            \end{enumerate}

        \subsection{Takeaway from the Interview Study}
        \label{interview discussion}

            In this section, we discuss the broader takeaways of interview findings \ref{interview findings} for nonprofit data and technology ecosystems, surfacing the need for an alternative approach to the design of an AI assistant that responds to the specific needs and vulnerabilities of IINGs. This discussion informs the subsequent steps of this study, specifically goals for design of the nonprofit AI assistant (\ref{design goals}). 

            The findings reveal how IINGs ,operating within politically sensitive contexts, experience systemic marginalization in the digital nonprofit ecosystem. Two interlocking forces shape this marginalization: exclusion from nonprofit-specific technologies and infrastructures, and the circulation of misinformation that distorts their missions and identities. These challenges are embedded in existing data ecosystems and risk intensifying as AI systems, particularly those powered by large language models (LLMs), become more tightly integrated into nonprofit workflows. This is especially consequential given the broad social impact of algorithmic tools\cite{alkhatib_street-level_2019, lipsky_street-level_2010,toyama_technology_2011}. Accordingly, AI systems, while promising, may deepen marginalization if built on top of non-inclusive infrastructures without a critical lens. This underscores the need for context-sensitive, accountable, and recourse-enabled AI design in the nonprofit sector \cite{dignazio_data_2020, darian_enacting_2023}.

            Simultaneously, IINGs must contend with misinformation about political affiliations and donation misuse; narratives that spread rapidly across messaging apps and social media, due to engagement-maximizing algorithms \cite{huszar_algorithmic_2022, vosoughi_spread_2018, flaxman_filter_2016}. 

            Together, exclusion and misinformation, combined with participants’ expressed need for AI support in day-to-day operations, point to the necessity of a system that not only offers essential technological services but also interrupts the fractured, metropole-centered design of existing nonprofit technologies \cite{lilly_irani_postcolonial_2010}. Through alternative design, an inclusive AI system must actively repair structural harms by supporting underrepresented groups’ digital presence, preempting misinformation, and bridging cultural, linguistic, and access divides. We translated these empirical insights into actionable design goals (Section~\ref{design goals}).

\section{Design Goals}     
        \label{design goals}
         We translated the interview findings into design goals that address the challenges faced by participants. In response to their desire for a tool to support the operational and productivity challenges (\ref{findings-infrastructure-shadow}-\ref{findings-tech-need}), we proposed the design of an AI assistant that is inclusive of informal nonprofit practices and groups with capacity gaps (\ref{findings-infrastructure-shadow}-\ref{findings-2nd-order-exclusion}); and acknowledges and takes action toward online misinformation (\ref{findings-infrastructure-shadow}-\ref{findings-2nd-order-misinfo}). Guided by principles of design justice \cite{costanza-chock_design_2020}, our approach recognizes that design should be led by those most affected by the technologies --here, immigrant nonprofit workers and their communities-- rather than imposed from outside. The proposed AI assistant serves two primary user groups: 1) \textit{donors and audience}, by assisting them in discovering trustworthy causes, making online donations, and tracking contributions and impact reports; and 2) \textit{nonprofit managers and volunteers}, by supporting them with fundraising, campaign and event planning, financial report generation, impact storytelling, volunteer coordination, workflow documentation, and organizational knowledge transfer. Unlike existing tools, which often exclude informal and politically marginalized efforts, this system is designed to directly address the socio-technical barriers surfaced in phase 1, translated into design goals to guide the development of the AI assistant:

    \begin{quote}
        \item \textbf{DG1) Support operation needs} Nonprofit workers described how organizational tasks consumed time that could otherwise be directed toward community service. They also mentioned their inconsistency in responding to donors' need to have a stable channel of donation, keep track of contributions receipts, and follow up on impact reports. Volunteers, likewise, lacked a clear pathway to explore causes within grassroots or informal groups. Thus, our first design objective is to provide a system that centralizes these needs into a single accessible platform.
        
        \textbf{DG2) Recognize informal charitable efforts} Many participants described how the lack of official nonprofit status deprived them of financial, managerial, and IT tools and services, reinforcing a cycle of disempowerment. Our second design objective is to create a system that validates and supports nonprofit efforts regardless of their legal registration status, enabling them to participate in broader nonprofit ecosystems.

        \textbf{DG3) Be accountable for misinformation} Participants shared that misinformation, often tied to political biases against the government of their country of origin, Iran, posed severe challenges to their work. Existing digital tools offered little to counteract these harms. Our third design objective is to develop a system that helps nonprofit practitioners proactively debunk, or \textit{prebunk}\cite{noauthor_prebunking_2025, barman_evaluating_2024, tay_comparison_2022} misinformation.
        
        \textbf{DG4) Bridge capacity gaps} Participants described how mainstream nonprofit platforms often presume high levels of English and digital proficiency, assumptions that effectively exclude large segments of their user base who are immigrants and/or seniors. These challenges are especially pronounced when the content, such as beneficiary narratives and reports, is handled by social workers in Iran in the Farsi language. Delivering services exclusively in English was described as unintuitive and misaligned with the linguistic and cultural realities.
    \end{quote}

        Designing this system not only meets practical needs but also supports everyday care practices that extend beyond formal organizations \cite{voida_infrastructures_2015}, and is sustained by a diverse range of individuals.

\section{Phase 2: Co-Design Session}
        The second phase of our study addressed the research question: \textit{“RQ2: How can we leverage design principles to support IINGs in harnessing emerging AI technologies while mitigating their adverse effects?”} In line with the design thinking process, the earlier interview study served as the first round of empathizing and defining, where we identified high-level pain points rooted in broader social infrastructures and specified preliminary design strategies \ref{design goals}. 
        
        Recognizing that design is inherently iterative, Phase 2 represented a subsequent cycle of empathizing and defining, this time directed toward eliciting more granular expectations from the AI assistant and translating previously derived high-level design goals into ideating concrete design features and interaction strategies. During this phase, we employed participatory design techniques, including expectation mapping, persona construction, and design ideation -methods that enabled participants to move from reflecting on lived challenges to envisioning actionable design possibilities.
        The methodological execution and empirical outcomes of this phase are detailed in the following section.

        \subsection{Methods} 
        \subsubsection{\textbf{Participant Recruitment}}
        Nine participants joined this phase of study, comprising 6 executive members and volunteers, and 3 frequent donors to charitable nonprofits. Seven participants were immigrants residing outside Iran, and 2 lived in Iran while collaborating with diaspora charities.
        \subsubsection{\textbf{Session Structure}} 
        In the co-design session, we conducted a 120-minute online study with 9 participants who were executive members and frequent donors of IINGs (\ref{methods overview}). The participants collectively envisioned expectations for a nonprofit AI assistant. The session unfolded in three main parts, detailed below, with each part lasting 20-30 minutes, followed by a group discussion recap. The participants used the Miro online collaborative platform~\cite{noauthor_miro_2025} to sketch their proposed designs and write their ideas.

           \textbf{(1) Expectation Mapping}: Participants engaged in an exercise designed to surface the everyday expectations and pain points that shape their charitable practices. Nonprofit executives were invited to identify routine tasks that consumed disproportionate amounts of time. They also discussed the kinds of intra- and inter-organizational coordination tasks that demanded the greatest emotional or physical energy. Donors, in turn, were asked to reflect on their interactions with charity groups: the expectations they typically had, information they wished to receive, and the factors that impacted their trust in organizations. This activity helped anchor the subsequent co-creating design exercise.
              
            \textbf{(2) Envisioning Personas}: Participants then engaged in a persona-building exercise to envision how an AI assistant might be experienced by diverse members of their communities. They constructed fictional characters that reflected a range of identities and backgrounds informed by participants' experience, such as variations in age, gender, technological literacy, and place of residence. Once the personas were articulated, participants envisioned how these individuals might interact with the AI system, either as donors or as volunteers. They considered the questions or challenges the personas could encounter during interaction. This activity encouraged participants to move beyond their own perspectives and anticipate the heterogeneous needs of their wider networks. 
            
            \textbf{(3) Co-creating Design Artifacts}: In the final activity, participants shifted from articulating needs and personas to creating concrete design features for the AI assistant. They were asked to imagine how the system could embody the design goals surfaced in phase 1 and core values articulated in expectation mapping and persona building. Working on a dedicated Miro board, participants created low-fidelity interface elements, conversational strategies, and interaction flows that could support these goals (Figure~\ref{fig: sketches}).

            \subsubsection{\textbf{Data Collection and Analysis}}
            The data collected in phase 2 included a full 120-minute session recording transcribed into 19 pages, participants’ notes and sketches on Miro collaboration boards, and the researcher’s notes during the meeting.
            We analyzed the session data using thematic analysis \cite{boyatzis_transforming_1998, braun_using_2006}. Following transcription and translation, we conducted open coding \cite{strauss_basics_1990}, allowing categories to emerge inductively from participants’ discussions and notes. Codes were iteratively compared and refined, and related categories were clustered into higher-level themes through affinity diagramming. ~Coding and disagreement resolution followed the consensus procedure described in~\ref{phase1_analysis}; Phase 1 themes served as sensitizing concepts, while coders remained open to categories without Phase 1 precedent, e.g., "volunteer autonomy". This process highlighted recurring concerns and suggestions of participants, which subsequently informed our Findings section.

        \subsection{Findings of the Co-design Session}
        \label{co-design findings}
        The co-design session revealed key expectations for the AI assistant through expectation mapping, identified potential barriers for diverse and intersectional user personas through persona envisioning, and generated concrete design elements aligned with design goals (\ref{design goals}). Together, these findings translate stakeholder needs and constraints into practical design directions tailored to IINGs.

        \subsubsection{\textbf{Findings from Expectation Mapping}}: In response to design goal DG1 (\ref{design goals}), donors reflected on their expectations of nonprofits, while the nonprofit executives focused on the recurring challenges they faced in practice. 

        \textit{Transparency, organizational clarity, and planning} emerged as recurring donor expectations in the discussions. Participants valued nonprofits with a clear mission and vision, well-structured planning, and campaigns focused on tangible, specific causes. They sought clarity about beneficiaries, especially criteria for selection and validation, and associated trust with specific, transparent communication and a coherent social media presence. Donors emphasized financial transparency, requesting detailed expense reports, \textit{quantitative statistics} on aid distribution, and infographic-style summaries of past work, alongside \textit{qualitative documentation} such as photos, stories, and frequent progress updates. Conversely, vague or overly general descriptions of activities (e.g., “helping those in need”) were cited as sources of distrust. Participant 1 articulates her expectation as below.

        \begin{quote}
            \textit{"I would like to see a clear mission and vision in the groups. They should know exactly what problem they want to focus on and communicate it well. For example, when they say "we focus on helping individuals with debt management", I can trust them. But giving vague phrases like "we're helping those in need" sounds distracted."} (P1)
        \end{quote}

        From the nonprofit executive perspective, \textit{communication and coordination} with both external audiences and internal team members were the hardest parts of their tasks, and they needed support from the AI assistant in these areas. 
        While donors placed exceptional emphasis on transparency and detailed reporting, nonprofit executives described these expectations as among the most labor- and time-intensive aspects of their work. Accounting and evaluation tasks were characterized as sensitive while fragmented, and requiring significant effort to transform into presentable outputs. Volunteers also reported devoting extensive effort to attracting and retaining supporters, as well as building trust, especially when contributions were indirect or non-monetary. 
        
        Beyond reporting, participants highlighted substantial coordination burdens, noting that team-based workflows, limited volunteer autonomy, and slow inter-organizational collaborations often created bottlenecks.
        In other terms, participants were asking the AI assistant to absorb part of the articulation work: to track task states, reduce the hand-offs between volunteers and across organizations, and maintain the shared records that coordination depends on~\cite{schmidt_taking_1992, suchman_supporting_1995}.
        
        A major structural gap was the absence of a systematic beneficiary identification and monitoring process, which led to confusion and repeated discussions about decision-making. Executives further underscored the challenges of recruiting and motivating volunteers over time. On the expectation mapping collaborative board on Miro Platform, P5 writes the tasks that are most challenging to their team:
                \begin{quote}
                    \textit{"-Persuading people to join as volunteers (supplying human resources)
                    \newline
                    -Preparing the detailed and sensitive accounting tasks, preparing reports, presenting them to supporters, and achieving their trust.
                    "} (P5 notes on Miro board)
                \end{quote}

        Taken together, these findings reveal a persistent gap: while donors expect transparency, detailed reporting, and clear monitoring, nonprofit workers struggle to meet these expectations under severe time, labor, and coordination constraints. This misalignment underscores the need for technological interventions to alleviate these burdens.
        
        \subsubsection{\textbf{Findings from Persona-Building}} In the "persona-building" activity, the envisioned personas included elderly supporters with limited digital literacy, refugees with language barriers, and individuals with disabilities, imagined to test the boundaries of system inclusivity. A notable finding was that, although participants’ charitable efforts often took place within their immediate immigrant networks, many reported collaborating with partners in Iran and even with donors from other countries. These interactions led them to imagine user personas in diverse geographic contexts, underscoring the need for a system capable of operating across linguistic and cultural boundaries.
        
        Across imagined users, two general categories of need emerged. First, \textit{accessibility and multilingual support} were critical. Diverse individuals highlighted the importance of these factors, including a 75-year-old Afghanistani woman living in Iran with minimal digital literacy, a second-generation Iranian immigrant eager to donate, and a blind Turkish-speaking student willing to mentor other students. For example, P1 explains her user persona as follows:

            \begin{quote}
                \textit{"She is a 25-year-old woman, lives in Tabriz, Iran, and speaks Turkish. Is blind and capable of working with software tools that help the blind. She is interested in volunteering and mentoring blind Turkish-speaking teenagers. She needs voice support when interacting with an AI tool. She will tell the AI assistant her skills, and the AI connects her to groups she can volunteer for."} (P1)
            \end{quote}
        
        Second, \textit{privacy and trust} consistently surfaced as essential design concerns. Donor personas, especially those with unstable immigrant status or those willing to donate to geographical regions with complex politics, demanded concrete evidence of safeguarding their identity and privacy when interacting with the system. P8 shares the concerns of her persona, who is a Chinese teenager who wants to send a donation to Middle East:
            \begin{quote}
                \textit{"He needs assurance that his conversation with the AI is not shared with the government, and his questions, identity, data, and transactions are safe and secure. "} (P8)
            \end{quote}

                 \begin{figure*}
            \centering
            \includegraphics[width=.75\linewidth]{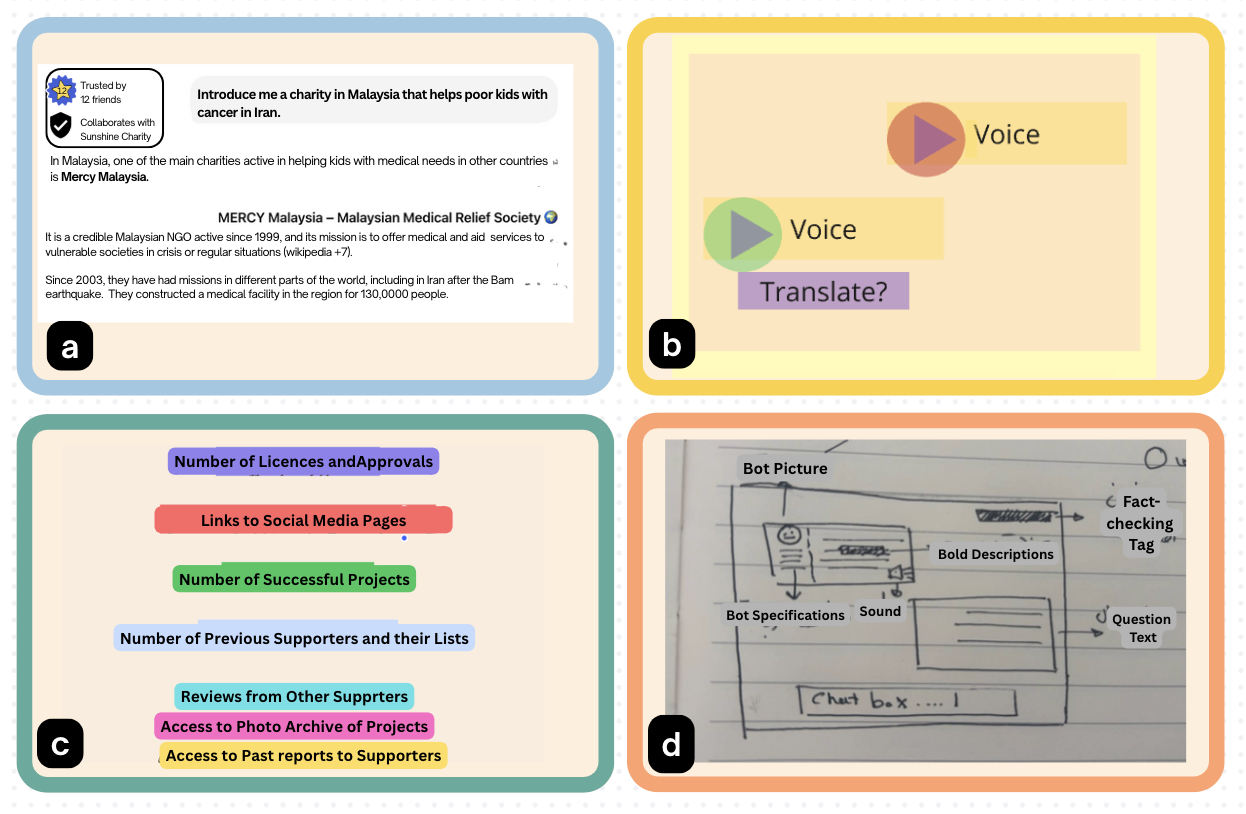}
            \caption{Examples of the participants' co-created element designs in the co-design session (notes are translated from Farsi). Figure (a) shows the concept of trust networks used as an alternative way for recognizing the legitimacy of non-formal groups in the AI tool. Figure (b) illustrates how participants imagined voice-based chats with simple elements, helping users with disability or low levels of digital proficiency. In figure (c), the participant has sketched simple buttons that can be placed at the end of chats with a conversational AI tool, each responding to frequent inquiries of donors. Finally, figure (d) shows expandable fact-checking tags to be placed in chats with AI, to offer more documentation about the claims and affiliations of nonprofits.}
            \label{fig: sketches}
        \end{figure*}

        \subsubsection{\textbf{Findings from the Co-creating Design Artifacts}} \label{co-designed artifacts} In the third part, participants proposed detailed design elements that directly responded to the design goals~\ref{design goals}, and the previous parts of the same co-design session. In response to DG2, addressing the legitimacy challenges faced by informal nonprofit groups, they proposed \textit{alternative validation mechanisms rooted in trust networks}. These included \textit{community verification tags}, that show endorsements from friends or well-established nonprofits (please refer to Figure~\ref{fig: sketches}-a to see the sketch of P2) as well as \textit{multi-criteria rating systems} that separately rate transparency, impact, and communication quality of the nonprofit groups (Figure~\ref{fig:AMINA-informality}-b)
        
        In response to concerns around misinformation (DG3), P5 proposed \textit{fact-checking features} (please see Figure \ref{fig: sketches}-d), such as expandable tags or buttons embedded in the AI interface. These elements would allow users to access additional documentation, such as financial reports, images, or videos, to verify the nonprofit’s activities and affiliations. 

        Finally, to bridge capacity gaps (DG4) and recognizing users with diverse levels of technical proficiency, language backgrounds, and abilities, P1 designed features such as \textit{voice-based interaction, translation, and simplified buttons} (please refer to Figure \ref{fig: sketches}-b).


\section{Phase 3: Designing AMINA: AI for Marginalized Immigrant Nonprofit Assistance}
    \label{phase3_prototype}
    Based on the insights from phases 1 and 2, we prototyped AMINA, the "AI for Marginalized Immigrant Nonprofit Assistance", on the Figma platform \cite{noauthor_figma_2025} and in iterative sessions. The acronym AMINA resembles the word "Amn", which means safe and trustworthy in most Middle Eastern languages. In addition to supporting the core functionalities needed in a nonprofit assistant tool, i.e., the core operational needs of nonprofit workers and donors (DG1), we incorporated the concerns and concrete design elements suggested in the co-design session into AMINA; specifically, the elements introduced in section \ref{co-designed artifacts}, such as community verification and rating of informal nonprofits as an alternative form of recognizing their legitimacy, fact-checking tags in response to misinformation, and voice-based communication and translation features to bridge access and capacity gaps (figure~\ref{fig: sketches}).
    
    This section presents AMINA’s system components as well as the user workflows that structure interaction with the assistant and guide the subsequent evaluation and feedback phase.
        \subsection{Components}
        The system architecture has two interoperable layers. The Explorer Agent is a stand-alone AI public assistant knowledgeable about the nonprofit ecosystem. It answers general queries (e.g., “Which groups support refugee education in Germany?” or “What are the political affiliations of Hope Foundation?”) using hybrid data sources, including formal directories, publicly scraped web and social media data, community-contributed metadata, and optional nonprofit-provided inputs. It also offers nonprofits a dashboard to manage their profiles, which is especially valuable for informal groups with limited digital presence.

        The Local Agents are nonprofit-specific instances of the Explorer Agent deployed on an organization’s website or social-media channels. They can integrate private or semi-private data, such as donation receipts, campaign reports, volunteer lists, and organizational documents, to answer questions like “How was my donation spent?” or “How can I join the next food drive?”



    \begin{figure}
  \centering
  \includegraphics[width=\textwidth]{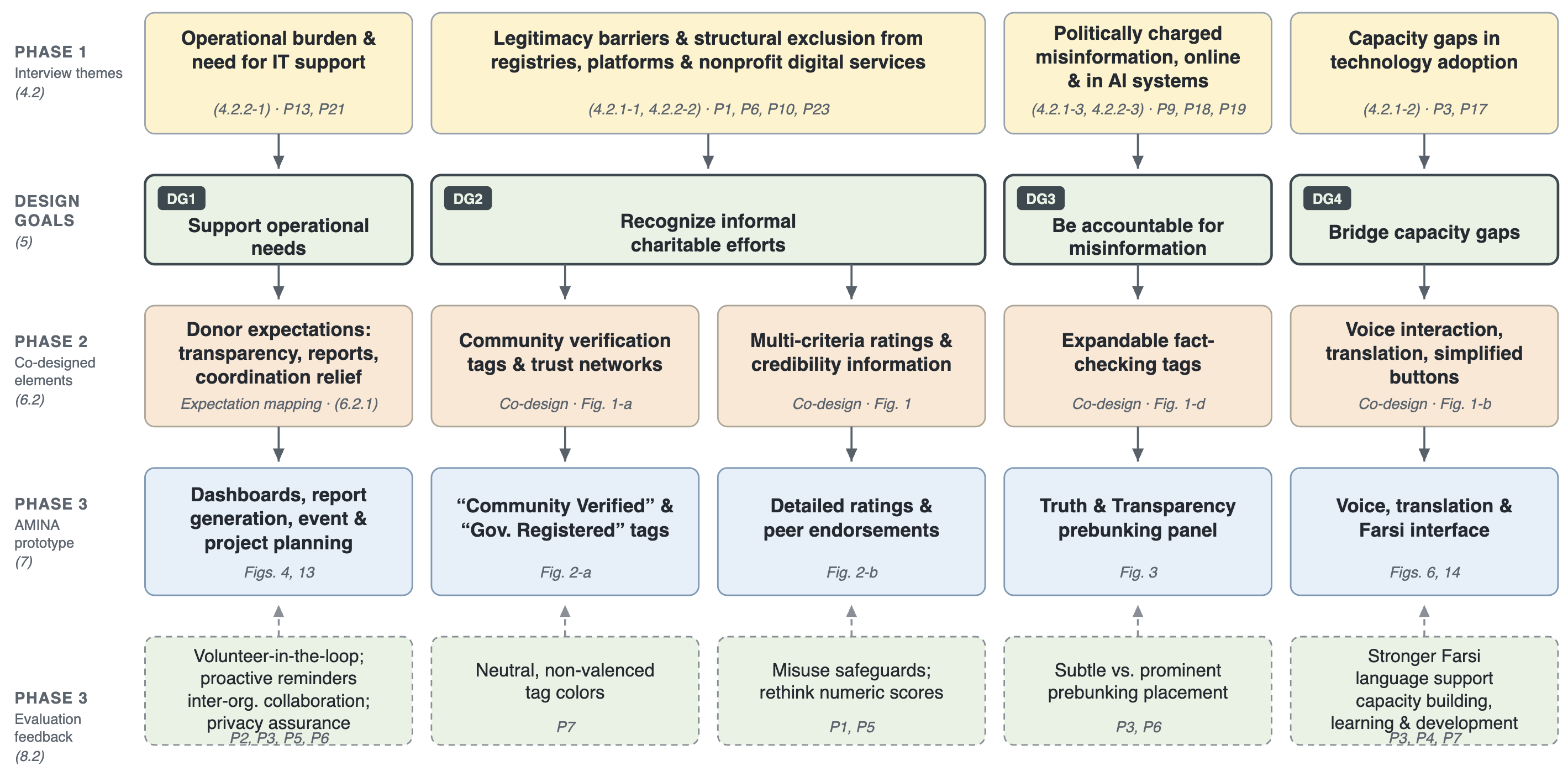}
  \caption{Design development of AMINA across the three study phases. Phase~1 interview findings (top) motivated the four design goals (\ref{design goals}); the Phase~2 co-design session translated these goals into concrete design elements (~\ref{co-designed artifacts}), which were implemented as AMINA prototype features (\ref{phase3_prototype}). Last row summarizes refinements surfaced in 
  Phase~3, user evaluation of AMINA (\ref{evaluation_findings}). Solid arrows indicate primary provenance; dashed arrows indicate evaluation feedback.}
  \Description{Flow diagram with five horizontal bands connected by downward arrows:
  Phase 1 interview themes map to four design goals DG1--DG4, which map to Phase 2
  co-designed elements, which map to AMINA prototype features; a bottom band of dashed
  boxes lists Phase 3 refinements feeding back into each feature, plus cross-cutting
  directions.}
  \label{fig:traceability}
\end{figure}

       \subsection{Workflow}
       \label{workflow}
        AMINA supports two primary user groups: (a) donors and the general public, and (b) nonprofit executives and volunteers. We outline typical workflows for each, which can be completed through communication with AMINA’s conversational interface (text or voice) or by manually interacting with graphical interface. While designed as an agentic conversational AI, the dual modality allows users to explore system capabilities and enables participants to visually inspect ideas and provide feedback.
        
    \textbf{(a) Donor and General Public Workflow.} 

        \textit{(1) Discover and Compare Nonprofits:} Users can ask the assistant to recommend various nonprofit groups, both registered and informal. The assistant provides an overview of each group’s mission, history, and current campaigns, drawing on its multilingual content.

        \textit{(2) Explore a Nonprofit Profile:} After selecting a nonprofit of interest, users can view a detailed profile that includes: (1) narrative and financial reports of past campaigns; (2) descriptions of ongoing campaigns; (3) the nonprofit's responses to common questions or misinformation; (4)indicators of credibility, such as community reviews and ratings, endorsement from friends and community, and/or governmental licenses.

        \textit{(3) Engage and Contribute:} Users can choose to: (1) make a donation through payment links integrated or via indirect links, (2) sign up as a volunteer via a guided form, (3) leave a public review.

        \textit{(4) Personal Dashboard:} Logged-in users have access to a personal dashboard that includes their history of interactions, favorite nonprofits, donation receipts, and history of chats with the AI assistant. This dashboard is designed to support long-term donor engagement.

        \textbf{(b) Nonprofit Organizer Workflow.} 
        
        \textit{(1) Nonprofit Dashboard and profile management} Executive members of nonprofits access a parallel dashboard interface that allows them to manage their presence on AMINA.
            
        \textit{(2) Event and Project Planning:} Organizers can plan upcoming events and projects using AI-assisted scheduling.
        
         \textit{(3) Report Generation:} The assistant encourages generating structured reports, both in quantitative form, like detailed financial reports and infographics, and qualitative, like impact narratives and beneficiary stories.
         
        \textit{(4) Credibility Work and Addressing Misinformation} The nonprofit organizers will be encouraged by AMINA to proactively share information about their political or religious affiliations (if any), beneficiaries, origin of monetary resources, and reports of expenditure to "pre-bunk" misinformation \cite{noauthor_prebunking_2025, barman_evaluating_2024, tay_comparison_2022}. 

        \textit{(5) Internal Documentation:} The organizers will record intra-group meeting notes, decision logs, and organizational knowledge and data to help sustain their activity.


        
We developed a high-fidelity interactive prototype of AMINA, incorporating the components described above to enable users to perform the specified workflow tasks. The final version of this prototype, employed in the evaluation interviews, is presented in Figures~\ref{fig:AMINA-informality}-~\ref{fig:AMINA-operations}, and in the appendix section.

While AMINA is presented to each user as a conversational assistant, its design goals are realized through features that support cooperation among users rather than individual productivity alone. First, AMINA functions as a coordination mechanism~\cite{schmidt_coordination_1996}: shared executive dashboards, project and event planning, meeting notes, and proactive reminders about task states that would otherwise live in individual volunteers' minds, protecting organizational continuity across the volunteer turnover. Second, its transparency artifacts like financial reports, impact narratives, and the prebunking panel operate as boundary objects~\cite{lee_boundary_2007,star_institutional_2025} between organizers and donors: representations produced within the nonprofit's workflow yet legible to outside supporters, allowing the two groups to coordinate trust. Third, AMINA is designed for human augmentation rather than substitution: this hands parts of the workflow to the executives, positioning the AI as a mediator within, not a replacement for, the group's relational fabric. Finally, the Explorer Agent's nonprofit directory extends the opportunity for collaboration beyond single groups, supporting inter-organizational knowledge sharing and resource pooling among otherwise isolated IINGs.

\begin{figure}
    \centering
    \includegraphics[width=0.7\linewidth]{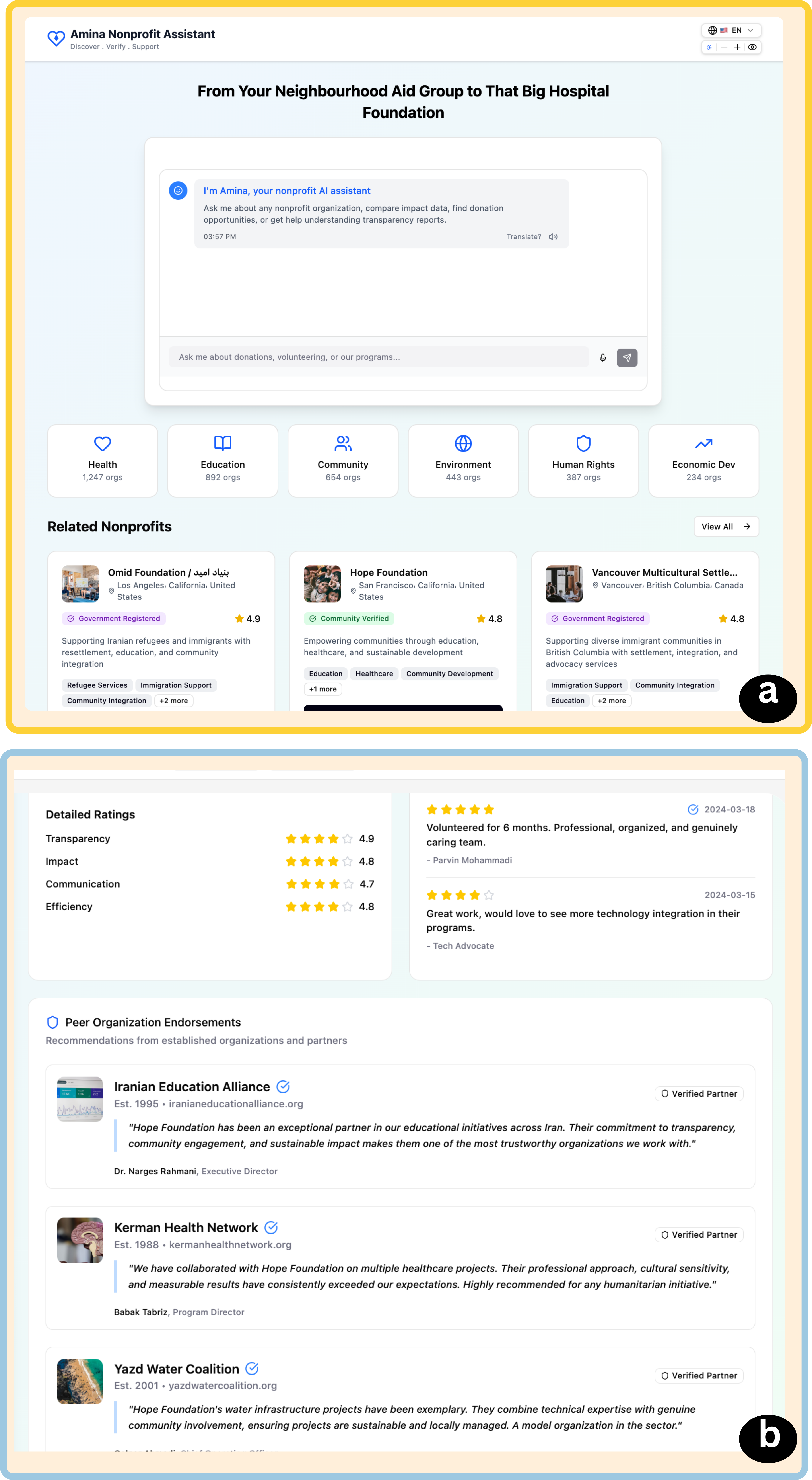}
    \caption{Example screenshots of AMINA prototype, indicating design artifacts that recognize informal nonprofit groups (DG2) by providing alternative ways of achieving legitimacy. (a) The landing page of the AMINA prototype, indicating the Explorer Agent, which presents the AI chat interface alongside a selection of nonprofit organizations; some organizations carry “Community Verified” tags, while others have “Government Registered” tags. (b) The "multi-criteria rating system and peer endorsement tab" that facilitates community evaluation of nonprofits.}
    \label{fig:AMINA-informality}
\end{figure}

\begin{figure}
    \centering
    \includegraphics[width=0.7\linewidth]{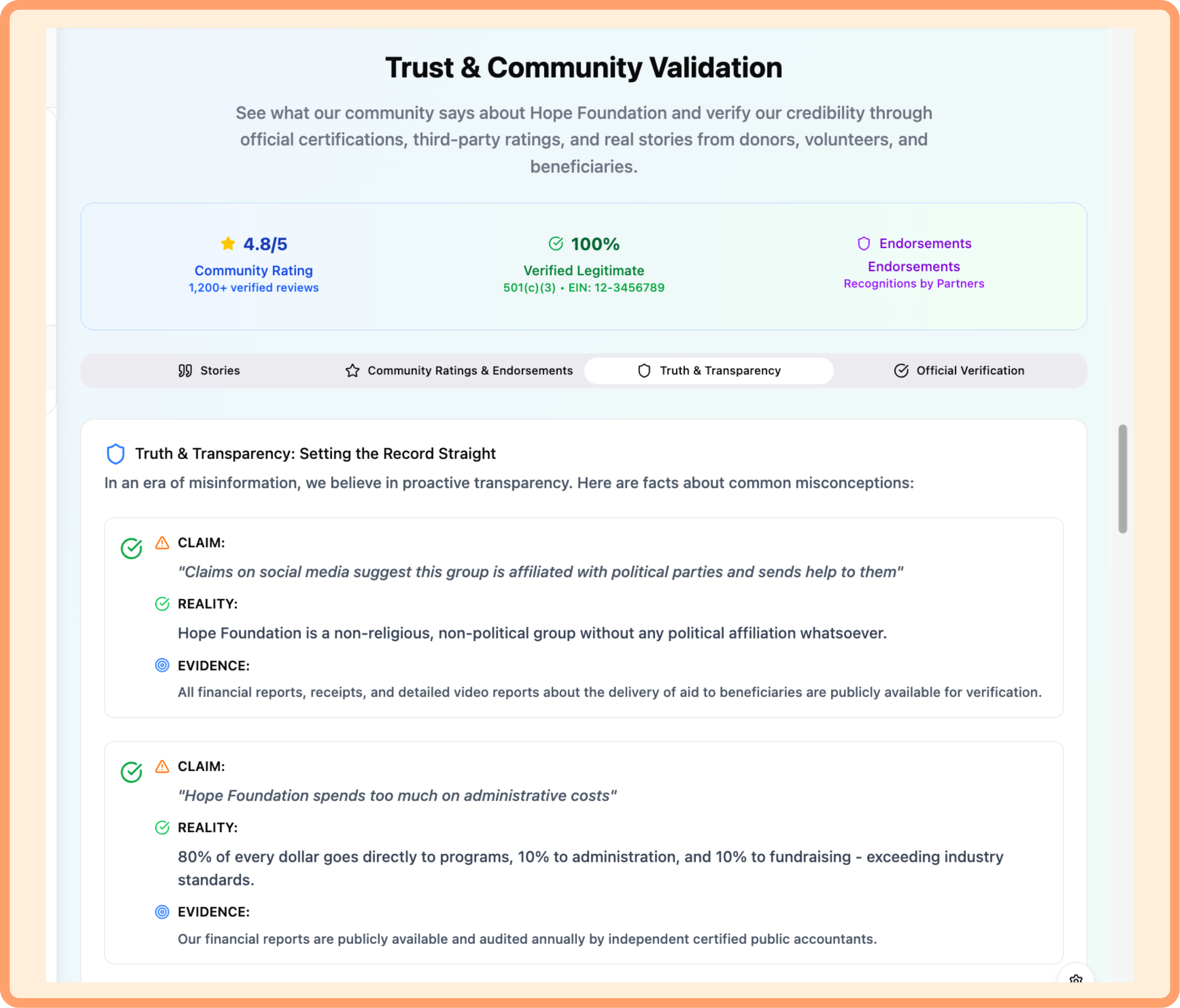}
    \caption{Example screenshot of AMINA prototype, indicating design artifacts that encourage nonprofits for proactive debunking of misinformation (DG3) by providing "fact-checking" buttons (in green, on the left), fields to include viral misinformation claims, true realities, and clickable sections to provide evidence.}
    \label{fig:AMINA-misinformation-large}
\end{figure}

\begin{figure}
    \centering
    \includegraphics[width=0.7\linewidth]{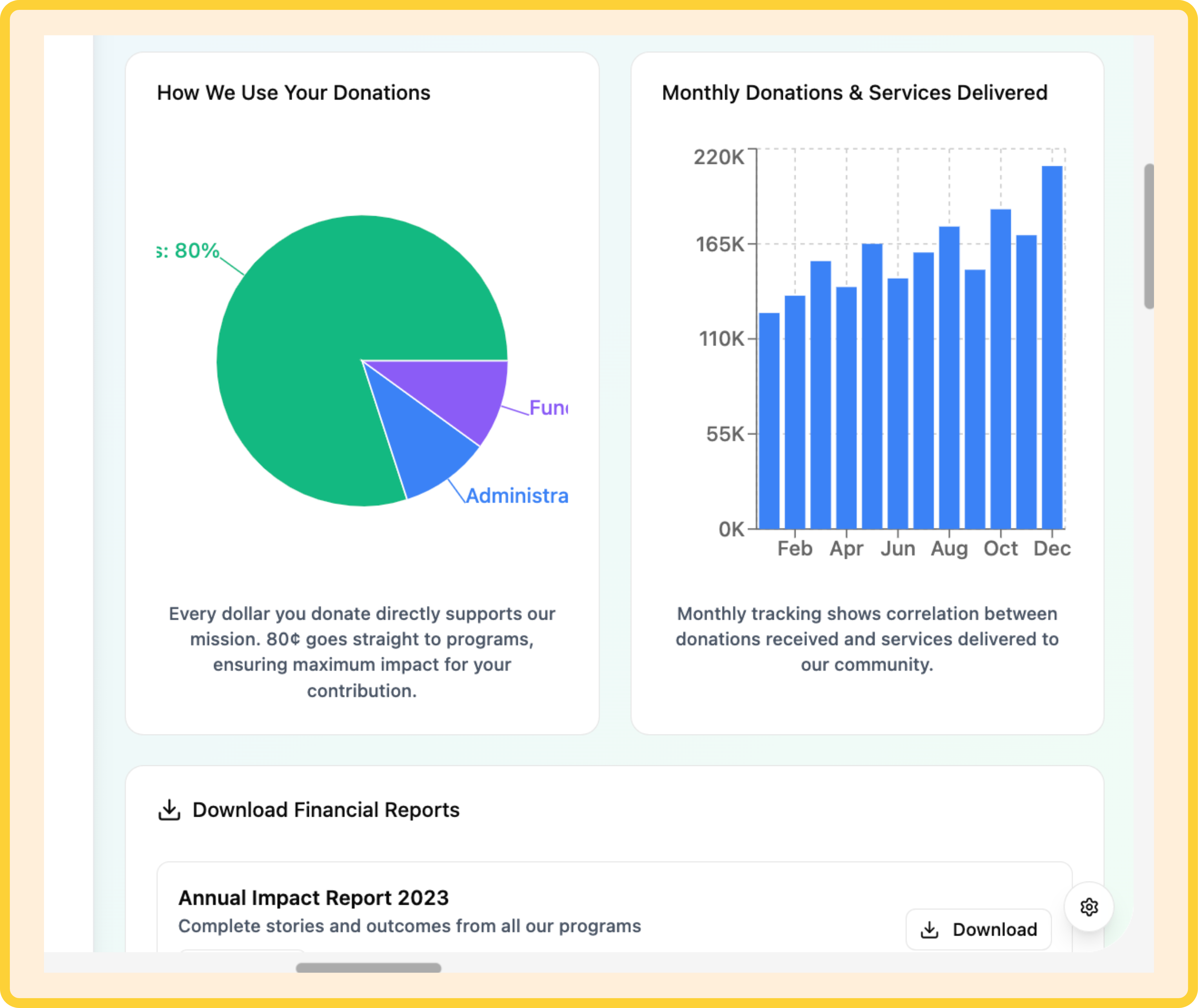}
    \caption{Example screenshot of AMINA prototype, indicating design artifacts that support operational needs of nonprofit groups (DG1). This figure indicates automatic AI-generated quantitative financial reports and an annual impact report. }
    \label{fig:AMINA-operations}
\end{figure}

\section{Users' Feedback on AMINA}
    Following prototyping, we conducted a formative evaluation of AMINA with seven practitioners. Our goal was not to measure task performance, but to assess whether the design concepts were legible, valued, and aligned with participants' practices, and to surface refinements for future iterations.
    
    \subsection{Methods}
    \subsubsection{\textbf{Participant Recruitment}}
            Seven participants joined the evaluation and feedback interviews; five identified as active members of IINGs, while two were frequent donors to such groups. Six participants resided in Global North countries, while one participant, a partner social worker of a group abroad, lived in Iran.
    \subsubsection{\textbf{Data Collection}} 
            We conducted online semi-structured interviews with participants, each session beginning with an introduction to the study and an overview of the tool. Participants were then asked for verbal consent to record both video and screen activity. We invited both user groups, donors and nonprofit volunteers, to complete both sets of workflow tasks, in order to elicit their perspectives not only on their own practices but also on those of the counterpart group with whom they routinely interact \ref{workflow}. Next, the researcher shared the link to AMINA via a published Figma site, asked participants to share their screens, and had them complete a sequence of tasks. They were encouraged to think aloud during the process, articulating their impressions and emotional responses. They were asked to take both an exploratory approach, eliciting opinions and feelings about the elements and concepts, and an evaluation approach, assessing their satisfaction with AMINA's usability. Ultimately, participants were invited to reflect on their overall experience and offer suggestions.
            The interface the participants interacted with was a high-fidelity but non-functional prototype: AI responses were scripted exemplars rather than live model outputs. Accordingly, the feedback reported below concerns design concepts, interaction flows, and interface framings, not the performance of an operational AI system. Claims regarding multilingual quality, fact-checking accuracy, and prebunking effectiveness should be read as design hypotheses that the evaluation helped refine, not as tested system capabilities.
            
        \subsubsection{\textbf{Data Analysis}}
        
         Each interview session lasted between 90 and 120 minutes (average 112 minutes), during which the researcher also took detailed observation notes capturing participant reactions, verbal cues, and design feedback. In total, this phase yielded over 13 hours of audio and screen recordings, transcribed and translated to English, which, along with session notes, resulted in 120 pages. We employed open coding \cite{strauss_basics_1990} and thematic analysis \cite{boyatzis_transforming_1998} on the session transcripts and notes. Two authors reviewed the transcripts, allowing codes to emerge inductively from the data. Following the initial round, the two coders compared codes, resolving discrepancies through discussion to reach consensus, as in ~\cite{mcdonald_reliability_2019}, and consulted session recordings when a participant's reaction was ambiguous. The consolidated codes were clustered into higher-order themes and cross-checked against the Phase 1 and Phase 2 thematic structures in a final integration meeting, then used to structure our findings.

    \subsection{Findings from Evaluations and Feedback}
    \label{evaluation_findings}



    Through their engagement with AMINA, participants offered feedback on the user interface and interaction flow, focusing on elements designed to operationalize the interview-derived design goals (~\ref{design goals}) and co-designed elements (\ref{co-designed artifacts}), particularly features supporting informality, misinformation, capacity gaps, and organizational operations. They also articulated broader design directions for AMINA’s future development. Below, we present the themes that emerged from our analysis. All names are pseudonyms.

        \begin{enumerate}

        \item {\textbf{Feedback on AMINA Elements Addressing Design Goals}}
            \begin{enumerate}
            \item {\textbf{Alternative Forms of Recognizing Informal Nonprofits}}
                    Participants reflected on the design elements derived from the co-designed artifacts \ref{co-designed artifacts} that introduced alternative mechanisms for recognizing informal nonprofit groups that lack access to formal legitimacy channels, i.e., government registration. These elements included the “community verified” tag, assigned through community input (Figures~\ref{fig: sketches}-a and \ref{fig:AMINA-informality}-a), as well as "multi-criteria rating systems" that allow users to evaluate different dimensions of nonprofits' performance such as transparency and impact (Figure~\ref{fig:AMINA-informality}-b).
                    
                    Participants responded positively to these mechanisms, emphasizing that they offer long-marginalized informal initiatives a pathway to visibility and digital support. As P3 noted:
                    
                    \begin{quote}
                    \textit{“I really appreciate the ‘community verified’ tag. Many people prefer to donate to community-based groups and may even distrust large government-registered charities, yet these are rarely acknowledged.”} (P3)
                    \end{quote}
                    
                    At the same time, participants expressed concerns regarding the presentation and potential unintended effects of these features. For instance, P7 cautioned that the visual design of the tags may inadvertently introduce bias:
                    
                    \begin{quote}
                    \textit{“Using green for ‘community verified’ suggests a positive judgment. Users might assume that groups with a purple ‘government registered’ tag, or no tag at all, are less trustworthy.”} (P7)
                    \end{quote}
                    
                    Participants also questioned the appropriateness of numeric scoring in the rating system. P5 expressed discomfort with quantifying humanitarian work:
                    
                    \begin{quote}
                    \textit{“Assigning numbers to a charity feels materialistic. These efforts are rooted in goodwill and compassion, and you cannot reduce the lives of beneficiaries to a numerical score.”} (P5)
                    \end{quote}
                    
                    Two participants extended these concerns to secondary negative impacts associated with legitimizing informal activities. As P1 warned:
                    
                    \begin{quote}
                    \textit{“It is important to recognize informal charity work, but you also need safeguards. Some might misuse this opportunity for scams or even money laundering. Maybe the AI itself can help detect suspicious behavior.”} (P1)
                    \end{quote}
                    
                    This feedback highlights both the promise and the risks of alternative legitimacy mechanisms. It underscores the need for careful design choices that prevent new forms of harm while supporting informal nonprofits in future iterations of the system.

            \item {\textbf{Prebunking Misinformation}}

                Participants evaluated AMINA’s approach to addressing misinformation through a dedicated panel in the prototype (Figure~\ref{fig:AMINA-misinformation-large}), which reflects the design goals \ref{design goals} and responds to the politically charged misinformation environment identified in the interview findings \ref{interview findings}. There was a strong consensus that the assistant must be equipped with accurate information and evidence so it can address users’ questions and clarify ambiguities arising from misinformation.

                Opinions diverged, however, on how explicitly this prebunking should appear in the interface. Five participants supported a prominent, proactive presentation of debunking content through a graphical tab. P6 expressed enthusiasm for this approach:
                
                \begin{quote}
                \textit{“This panel is very useful, especially the part that provides evidence for what is actually happening. It should appear even on AMINA’s first page.”} (P6)
                \end{quote}
                
                In contrast, two participants cautioned that explicitly placing a misinformation panel could unintentionally undermine trust. As P3 explained:
                
                \begin{quote}
                \textit{“Displaying a misinformation tab so prominently might plant seeds of doubt. Users may wonder why something must be wrong that the group is emphasizing on this matter.”} (P3)
                \end{quote}
                
                These participants recommended either making the panel more subtle or having the conversational AI agent present accurate information when users request clarification. Together, this feedback highlights the design tension between proactive prebunking and the risk of inadvertently signaling distrust.

            \item{\textbf{Performance in Farsi Language}}
            Another recurrent concern was AMINA’s performance in non-English contexts. While many donors and volunteers in immigrant communities have some English proficiency, collaborators in Iran often do not. Participants stressed the need for an improved Farsi interface and conversation that is both viable and appropriately toned, as P7 explained:
            
            \begin{quote}
            \textit{“Ms. Azimi is the direct contact in Iran with the kids we support. She sends me many images and video clips showing how donations are used. I wonder if she could upload them directly to this tool and have the AI generate a report. But she would need better Farsi support. ChatGPT does relatively well [in Farsi]. Will you build this tool with GPT?"} (P7)
            \end{quote}

            \item {\textbf{Supporting Organizational Operations}}
            Participants also reflected on AMINA’s performance in supporting core nonprofit operations. While they valued its capabilities for fundraising, report generation, and project and event management, they expressed concerns about how these features were presented. They identified a tension between text-dense interfaces and the needs of users who preferred minimal or voice-only interactions, noting that key organizational information was sometimes hidden behind tabs. To address this, P1 suggested a customizable interface that enables flexible switching between detailed views and simplified or voice-based modes. 
            \end{enumerate}    
        
        \item{\textbf{Suggestions on Design Strategies}}

            \begin{enumerate}

            \item {\textbf{Volunteer in the Loop}}
            Participants mentioned the need for AMINA to have human-like interaction that feels neither overly robotic nor excessively enthusiastic. They especially emphasized on volunteer-in-the-loop mechanisms to preserve trust, which is central to donors’ willingness to contribute; they stressed the need for reassurance that donations are actively monitored and used responsibly. As P5 reflected on their organization’s fully autonomous website:

                \begin{quote}
                    \textit{"A friend once told me she had submitted a volunteer sign-up form to our website several months ago, but never heard back from us. That’s because the website doesn’t give us proper notification, and we don’t check it often either. We don’t want that to happen with this AI tool."} (P5)
                \end{quote}

            She stressed that when the AI assistant cannot fulfill a request, it should promptly escalate to a human nonprofit volunteer; an approach we know as “human-in-the-loop” \cite{amershi_power_2014, wang_humans_2019}, which is essential for maintaining donor trust and avoiding perceptions of abandonment by the nonprofit group. Participants also emphasized the importance of sustaining human connection among volunteers and between donors and volunteers, an aspect that AI cannot replace and should not undermine.

            \item{\textbf{Support for Proactivity}}
            \label{proactivity}
                A recurring theme was the need for AMINA to act not only as a reactive tool but as a proactive agent. Participants emphasized proactive strategies across multiple dimensions, in addition to previously discussed countering of misinformation: practicing transparency and organizational sustainability, and safeguarding trust. 
                
                Regarding transparency and organizational sustainability, participants valued prompts that encouraged nonprofits to be more intentional in uploading expense receipts, photos, meeting minutes, and coordination summaries, tasks often neglected in under-resourced contexts. As P3 explained:
                
                \begin{quote}
                \textit{"It would be great if the system reminds us of uploading documents. We always talk in our team about sharing more photos [of the project impacts], and even sharing accounting details and summaries of coordination within ourselves."} (P3)
                \end{quote}
                
                Beyond task reminders, participants highlighted the importance of proactively communicating data privacy safeguards. This was especially salient in politically sensitive contexts, where activists feared surveillance. For instance, P6 questioned whether marginalized groups would feel safe entering sensitive beneficiary data or seeking legal guidance through the system.
                
                \begin{quote}
                \textit{"I'm thinking when the [volunteer] user wants to use this system to enter their beneficiary data for report generation or ask legal advice from the AI system, do they feel safe enough to do so? For example, imagine the Afghanistani immigrants who are at risk of being deported from Iran\footnote{Afghan immigrants in Iran, many of whom are refugees or undocumented, face long-standing structural discrimination, legal precarity, and episodes of forced deportation. These conditions have shaped their cautious relationship with state systems and digital infrastructures.} They need assurance."} (P6)
                \end{quote}
                
                Together, these insights highlight that AMINA must be proactive not only in supporting operational tasks but also in carefully balancing transparency, misinformation strategies, and privacy assurances to foster trust in this context.

                \item{\textbf{Support for Inter-organizational Collaboration}}
               Many IINGs operate in isolation in different communities, leading individuals to start new nonprofits from scratch and face infrastructural barriers that others have already overcome with hardship. Meanwhile, already established nonprofits struggle with limited human resources and need volunteer support. Participants suggested that AMINA could help by facilitating collaboration and knowledge-sharing, enabling more efficient use of time, energy, and expertise across different IINGs. Participant 2 suggested AMINA as a network hub for resource-sharing and collaboration, not just an individual assistant.

                \begin{quote}
                \textit{"The key outcome of this tool should be a reference network of every charity within communities of different locations, so that they can use each other's experience, and collaborate, instead of repeating each other's mistakes."} (P2)
                \end{quote}

                \item{\textbf{Facilitate Capacity Building, Learning and Development}}
                Participants recommended onboarding support, such as guided walkthroughs, to assist volunteers with limited digital confidence. Beyond this basic onboarding, they envisioned AMINA proactively identifying emerging or disconnected initiatives and reaching out to them, rather than waiting to be discovered by them. Such proactivity was viewed as essential for overcoming cultural, linguistic, and digital proficiency barriers. As P4 explained:

                    \begin{quote}
                        \textit{"Can you add onboarding tutorials like apps offer at first use? They would help me understand what features exist and what to expect, especially since AI might have some features that I might not know at all."} (P4)
                    \end{quote}
                
                Participants also saw value in AMINA for long-term organizational learning and development. They proposed incorporating educational modules on topics such as digital marketing and management for nonprofits to build capacity, particularly for newcomers to the nonprofit ecosystem. As P3 explained:

                    \begin{quote}
                        \textit{"A friend of mine once suggested taking Six Sigma~\footnote{Six Sigma courses teach process improvement and problem-solving methodologies to help organizations reduce defects, increase efficiency, and enhance quality.} courses to improve my managerial skills. I've always wanted to follow up on that. It would be nice to have courses like this in the tool, to keep all aspects of our work in one place."} (P3)
                    \end{quote}

         In sum, participants envisioned AMINA as an empathetic, contextually grounded, proactive assistant that anticipates needs, encourages transparency, assures privacy and safety, catalyzes collaboration, and supports capacity building. 
    
            \end{enumerate}

        \end{enumerate}

\section{Discussions}
    The findings from the interviews, co-design session, and evaluation and feedback sessions call for discussions around the design of inclusive and accountable AI technologies for nonprofit efforts in sensitive political backgrounds. Below, we situate the findings of this study within the broader field of HCI, highlighting concepts of immigration, informality, proactivity in design, and misinformation.
    
    This study extends the focus of migration-HCI from individual concerns of migrants to their collective practices~\cite{sabie_decade_2022, sabie_migration_2021, williams_hci_2024}. By examining socio-technical practices of IING activists, this work joins prior scholarly work in HCI on immigrants' sense of belonging and settlement~\cite{ ana_maria_bustamante_duarte_participatory_2018, hsiao_how_2023, lei_unpacking_2024}, bureaucratic challenges~\cite{coles-kemp_accessing_2019, dina_sabie_moving_2019}, concerns about data and identity protection \cite{Abu-Salma_diverse_2023, Steinbrink_digital_2021}, and privacy, safety and trust in immigrant communities \cite{hsiao_how_2023, kruger_what_2021}. 

    Our study also contributes to work on proactivity and nudging in HCI by demonstrating their importance in low-resource and politically sensitive nonprofit contexts. We use the term 'nudging' to denote subtle design interventions that guide user behaviors through reminders and feedback, consistent with the persuasive and reflective design literature \cite{thaler_nudge_2009, caraban_23_2019, zhang_nudge_2021}. We extend the conversation on nudging and proactivity~\cite{caraban_23_2019, zhang_nudge_2021, monge_roffarello_nudging_2023, m_dicosola_iii_nudging_2022} by showing how proactivity sustains volunteer-run groups facing legal vulnerability, misinformation, and resource scarcity. Key domains of proactivity influence, as mentioned by participants in \ref{proactivity}, include prebunking misinformation \cite{biselli_mitigating_2025}, documenting organizational knowledge \cite{pejovic_anticipatory_2015}, and offering timely privacy and safety nudges to users. Such interventions must avoid digital noise \cite{jarrahi_mindful_2023, zargham_understanding_2022, oh_better_2024} and respect user agency in sensitive sociopolitical contexts \cite{zargham_understanding_2022, oh_better_2024, clausen_exploring_2023, jamshed_designing_2025}. In doing so, our work expands proactive HCI beyond commercial and health domains \cite{ahire_workfit_2024, mehrotra_my_2016, pejovic_anticipatory_2015} to the nonprofit sector.

    Beyond immigration and proactivity, this work extends the scholarship on informality and misinformation, discussed below.

    \subsection{Informal Practices in HCI}
        Our study highlights informality as central to the everyday practices of IINGs under geopolitical constraint. From a design justice perspective, these informal infrastructures constitute community innovations that emerge in response to geopolitical exclusion, sanctions, and platform-level erasure. Rather than treating them as gaps requiring formalization, design justice encourages designers to begin with “what is already working at the community level,” recognizing informality as an asset rather than a deficit.

        The findings about informal practices of IINGs align with long-standing HCI scholarship that demonstrates how situated action and invisible work sustain collective activity when formal systems fall short. Suchman’s account of situated improvisation \cite{Suchman_Plans_1987}, and Star and Ruhleder’s concept of invisible infrastructural labor \cite{star_steps_1996, star_ethnography_1999, star_layers_1999} show that much cooperative work relies on tacit, adaptive, and often unrecognized forms of coordination. Likewise, the articulation of the social–technical gap~\cite{ackerman_intellectual_2000} and studies of informality in healthcare  \cite{Hardstone_Supporting_2004} emphasize the inevitability of informal workarounds in socio-technical systems. Our participants’ practices, ranging from banking workarounds to reliance on messaging groups, resonate with this lineage, framing informality not as a flaw but as the very mechanism by which nonprofit work persists amid fractured infrastructures. 

        Design justice grounds these informal practices in structural conditions and highlights how communities adapt within matrices of domination, including sanctions and restricted financial infrastructures that shape the forms of organizational action that are possible. In this line, our findings extend recent studies of informality in finance and community infrastructures, including ethnographies of informal electronics markets \cite{chandra_informality_2017}, analyses of sanctions-era community finance practices \cite{rohanifar_money_2021}, and prototypes of community-owned remittance tools \cite{Rohanifar_Kabootar_2022}. These works foreground the necessity of informal arrangements in contexts of political exclusion. For immigrant-led nonprofits tied to sanctioned regions, informality enables continuity and trust where registries, banking systems, and official data sources are inaccessible. In this setting, informality functions as a survival strategy to systemic exclusion \cite{karusala_future_2021}. Building on postcolonial critiques of urban informality \cite{Varley_Postcolonialising_2013}, we also understand these practices not merely as functional adaptations but as politically and culturally situated acts of resistance that challenge formal/informal binaries. By prototyping and studying an AI assistant in this context, we contribute a novel perspective: AI can be designed to scaffold informality rather than suppress it, aligning with design justice principles that call for technologies to support, rather than overwrite, community-defined practices.

        While recognizing the importance of supporting informal nonprofits, it is also necessary to acknowledge potential risks which confronts us with a genuine trade-off: the mechanisms that extend legitimacy to excluded informal groups can extend it to bad actors, namely the scams and money laundering our participants themselves warned of. Verification thresholds are thus choices about who bears risk: strict gatekeeping reproduces the exclusion documented in Phase 1 and pushes verification labor back onto volunteers~\cite{tanaka_legitimacy_2016}, while open inclusion shifts risk onto donors and onto the credibility of the informal sector as a whole, where fraud is a documented if understudied problem~\cite{zenone_fraud_2019}. There are design directions that can soften, though never dissolve, this tension, including graduated trust, in which platform privileges unlock stepwise as a group builds a verifiable track record. For example, profile visibility first, donation processing and higher contribution limits only after sustained community endorsement(Fig~\ref{fig:AMINA-informality}-b), and AI-assisted anomaly detection routed to community reviewers. Designing well for this trade-off requires understanding of how vouching and reputational stakes function within community trust networks and whether they deter misuse~\cite{besley_group_1995}; which legitimacy signals community members themselves find credible, as against those platforms privilege~\cite{hsiao_how_2023,tanaka_legitimacy_2016,ma_self-disclosure_2017}; the tactics and prevalence of bad actors~\cite{zenone_fraud_2019,siering_detecting_2016} in informal charitable channels; where responsibility should rest when community verification fails~\cite{nissenbaum_accountability_1996}; and how verification interacts with the safety needs of groups that depend on invisibility~\cite{jensen_alternative_2015,asad_illegitimate_2015,star_layers_1999}
    
        This framing positions informal practices as working community expertise that design must honor, sustain, and build upon, while remaining vigilant about potential risks.
    
        \subsection{HCI and Misinformation}
        
        Our study extends the growing body of HCI scholarship on misinformation interventions by foregrounding the challenges of IINGs, who often face politics-related misinformation. From a design justice perspective, misinformation targeting this population cannot be treated as a neutral information deficit; it is entangled with geopolitically charged narratives that cast these groups as risky or illegitimate. These structural forces shape not only the types of misinformation that circulate but also the credibility and impact of IINGs within diasporic communities. This view echoes arguments that misinformation cannot be treated as an isolated cognitive problem but must be understood through the socio-technical infrastructures and power asymmetries. \cite{marres_why_2018,davies_sts_2022,bounegru_field_2018}.

        Prior design strategies in addressing online misinformation have emphasized frictions that urge users to reconsider the content before sharing online \cite{jahanbakhsh_exploring_2021,greenberg_popups_2024}, and highlighted the importance of temporal alignment between exposure and intervention \cite{wilner_its_2023}. In continuation, our findings show that prebunking \cite{noauthor_prebunking_2025, barman_evaluating_2024, tay_comparison_2022} is particularly crucial in nonprofit contexts, where misinformation can directly undermine donor trust. Yet, the ways prebunking is surfaced in an interface matter significantly, as most participants advocated for an explicit, always-visible prebunking panel, arguing that clear evidence and corrective explanations should be immediately accessible. Others preferred more subtle approaches. These contrasting perspectives highlight the importance of prebunking designs that are sensitive to community trust dynamics and avoid signaling doubt. In this way, our work joins and extends research on prebunking as a more durable strategy for building resilience to manipulation, compared to reactive fact-checking~\cite{roozenbeek_fake_2019,tay_comparison_2022,barman_evaluating_2024,biselli_mitigating_2025}. For the population of this study, prebunking serves as a corrective to the narrative inequities embedded in socio-technical systems that routinely misrecognize or problematize their activities.

        By situating prebunking within the everyday practices of immigrants, we add to frameworks that systematize nudge techniques for combating misinformation \cite{konstantinou_nudging_2023,konstantinou_behavior_2025} and studies of community-driven fact-checking interventions~\cite{chuai_did_2024,amiri_besheli_combining_2025}. 

    \subsection{AI as a Collaborative Partner in Nonprofit Work}
        Our findings position AI assistants in nonprofit contexts not as automation of individual tasks but as participants in ongoing cooperative work. First, AMINA operates as a coordination mechanism that absorbs articulation work~\cite{suchman_supporting_1995, schmidt_taking_1992,schmidt_coordination_1996}. The labor our participants found most draining, including financial accounting, assembling reports, and preserving organizational memory across volunteer turnover, is precisely the meshing work that CSCW has long shown to be essential yet invisible~\cite{star_layers_1999}. Delegating portions of it to an AI assistant does not merely save time; it changes which cooperative arrangements remain viable for groups whose scarcest resource is volunteer labor.

        Second, AMINA mediates between communities of practice with asymmetric needs. Donors demand transparency; organizers experience producing it as a burden~\ref{co-design findings}. AMINA's reports, ratings, and prebunking panels act as boundary-negotiating artifacts~\cite{lee_boundary_2007}: representations through which the two groups work out what counts as sufficient evidence of legitimacy. This is a function that generic nonprofit platforms currently do not offer for informal groups. Lee's insight~\cite{lee_boundary_2007} that such artifacts are contested and continually renegotiated, rather than standardized and stable, is borne out directly in our Phase 3 data: the disagreement over how prominently prebunking should appear~\ref{evaluation_findings} is a live negotiation over how the artifact should present itself to different audiences, rather than one-size-fits-all transparency displays.

        Third, our volunteer-in-the-loop findings qualify the emerging vision of AI as a teammate~\cite{seeber_machines_2020,wang_human-ai_2019, he_ai_2024}. In commercial and data-work settings, human-AI collaboration research asks how machine teammates can take on ever more capable roles~\cite{seeber_machines_2020}; our participants inverted the question, asking how the AI can route the right moments back to humans. In trust-reliant, politically fraught charitable work, the assistant's collaborative value lies partly in knowing what not to handle: privacy-sensitive questions and moments where a donor needs to feel that a person stands behind the organization~\ref{evaluation_findings}. AI as a collaborative partner, in this setting, means an actor that facilitates coordination, translates across stakeholder boundaries, and deliberately re-centers human relationships. This is a design stance consistent with mixed-initiative principles~\cite{horvitz_principles_1999,amershi_power_2014} and with our design-justice commitment that technology scaffolds rather than overwrites community practice.
\section{Limitations}
This study has several limitations that constrain its generalizability. Recruitment drew primarily from authors’ networks in North America and Europe, leaving Iranian diaspora communities in East and South Asia, Turkey, Arab countries, and Australia underrepresented. The sample also focused exclusively on immigrants of Iranian origin, although other diasporic populations engage in distinct forms of transnational philanthropy, such as American expatriate nonprofits in South Asia. Moreover, we centered donors and volunteer organizers, excluding perspectives from end-beneficiaries, state actors, and intermediary nonprofits in Iran. Our recruitment strategy also introduces the possibility of network-based sampling bias. Because participants were recruited through the first author's channels and participant referrals, individuals connected to these networks may be more likely to be represented than those outside them. This may have shaped the range of perspectives regarding political sensitivities, misinformation, and legitimacy challenges surrounding Iranian immigrant nonprofit activity. At the same time, the participants brought varied organizational roles, geographic contexts, and experiences with broad-scale community work, both in charitable nonprofit work and in non-charitable settings, allowing the study to capture multiple perspectives on these phenomena.
As with all qualitative research, sampling biases are inherent, and our aim was not statistical generalization but the production of situated knowledge \cite{sedgwick_convenience_2013, lofland_analyzing_2022}. We therefore view the findings as situated accounts of recurring challenges and design needs within the studied community rather than as statistically representative estimates of the broader Iranian diaspora.

Despite these constraints, the results illuminate socio-technical challenges shared by immigrant nonprofits from other politically complex contexts and demonstrate the feasibility of inclusive AI tools under similar conditions. Finally, the AMINA prototype, implemented as a high-fidelity Figma mock-up without backend integration, enabled rich design feedback but means our findings speak to the perceived value and legibility of AMINA's design concepts rather than to the efficacy of its AI components, which requires a functional deployment to assess.

\section{Future of the Work and Design Implications}
Building on this study’s findings, future design and research on immigrant nonprofit efforts should pursue several directions. First, AMINA should strengthen proactive reminders that support transparency, organizational sustainability, privacy assurance, and the prebunking of misinformation, while ensuring that AI augments rather than replaces volunteer labor in the cooperative setting. Second, given the importance of empathy in nonprofit work, future iterations should incorporate human-like qualities, such as empathetic tone, culturally resonant communication, and multilingual fluency, to maintain emotional and cultural coherence. Third, AMINA’s capabilities could be extended to directly support the end-beneficiaries through resource introductions or empowerment programs. Finally, future work should examine AMINA’s applicability to non-Iranian immigrant groups and other under-resourced nonprofits to assess transferability and refine the system’s design.
\section{Conclusion}
This paper examined how IINGs navigate legitimacy barriers, fractured infrastructures, and politically charged misinformation under complex geopolitical conditions. Our study traced these challenges and explored how emerging AI technologies might ease them.

Our contributions are threefold: (1) We presented an empirical understanding of how socio-technical exclusions and misinformation shape immigrant nonprofit work; (2) We designed and evaluated an AI assistant prototype that addresses nonprofit and donor needs, and fosters nonprofit credibility and transparency without excluding informal nonprofit efforts; and (3) presented implications for justice-oriented AI design that positions immigrant-based nonprofits as critical sites of care work. Together, these contributions expand HCI scholarship on immigration, informality, proactivity,  misinformation, and nonprofit technology, and chart pathways for designing AI that strengthens, rather than replaces, the human connections at the core of nonprofit practice.

\bibliographystyle{ACM-Reference-Format}
\bibliography{references}

\clearpage

\appendix
\section{Appendix: Visuals of AMINA Interface Components }
In this appendix section, we present screenshots from the Figma-based prototype of AMINA, the AI for Marginalized Immigrant Nonprofit Assistance. Each figure below illustrates a major component of the prototype interface. Please note that only the most salient elements of each page of prototype are shown here, rather than a full view of pages. Please refer to the captions to learn about the functionality of each design element and its role in AMINA prototype.

\begin{figure}[h]
    \centering

    \begin{minipage}{0.75\textwidth}
        \includegraphics[width=\linewidth]{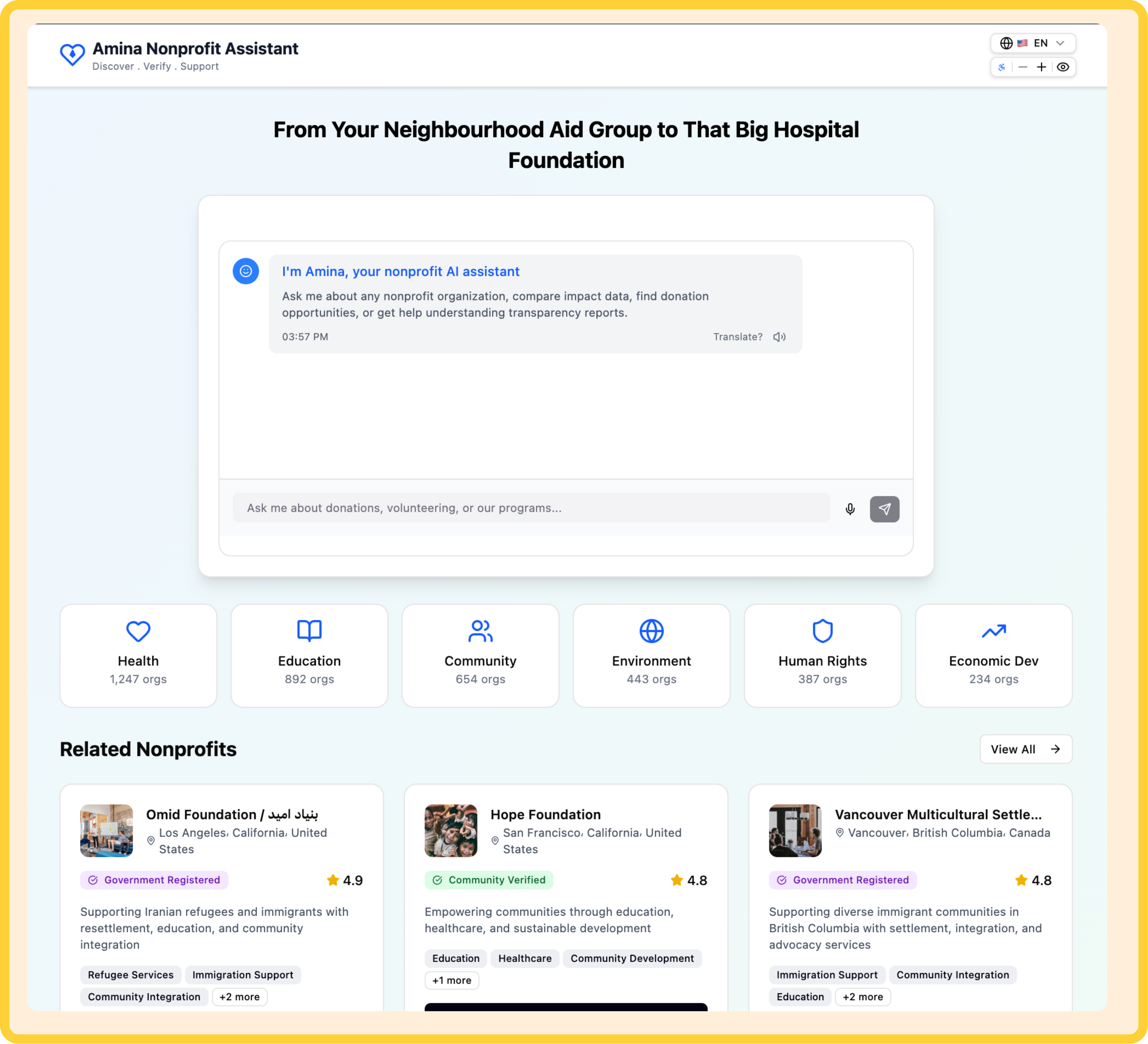}
    \end{minipage}
    \hfill
    \begin{minipage}{0.4\textwidth}
        \caption{The landing page of AMINA, and the Explorer Agent, which presents the AI chat interface alongside a selection of nonprofit organizations; some organizations carry “Community Verified”, while others have “Government Registered” tags.}
        \label{fig:AMINA-a.png}
    \end{minipage}
\end{figure}

\begin{figure}[h]
    \centering

    \begin{minipage}{0.6\textwidth}
        \includegraphics[width=\linewidth]{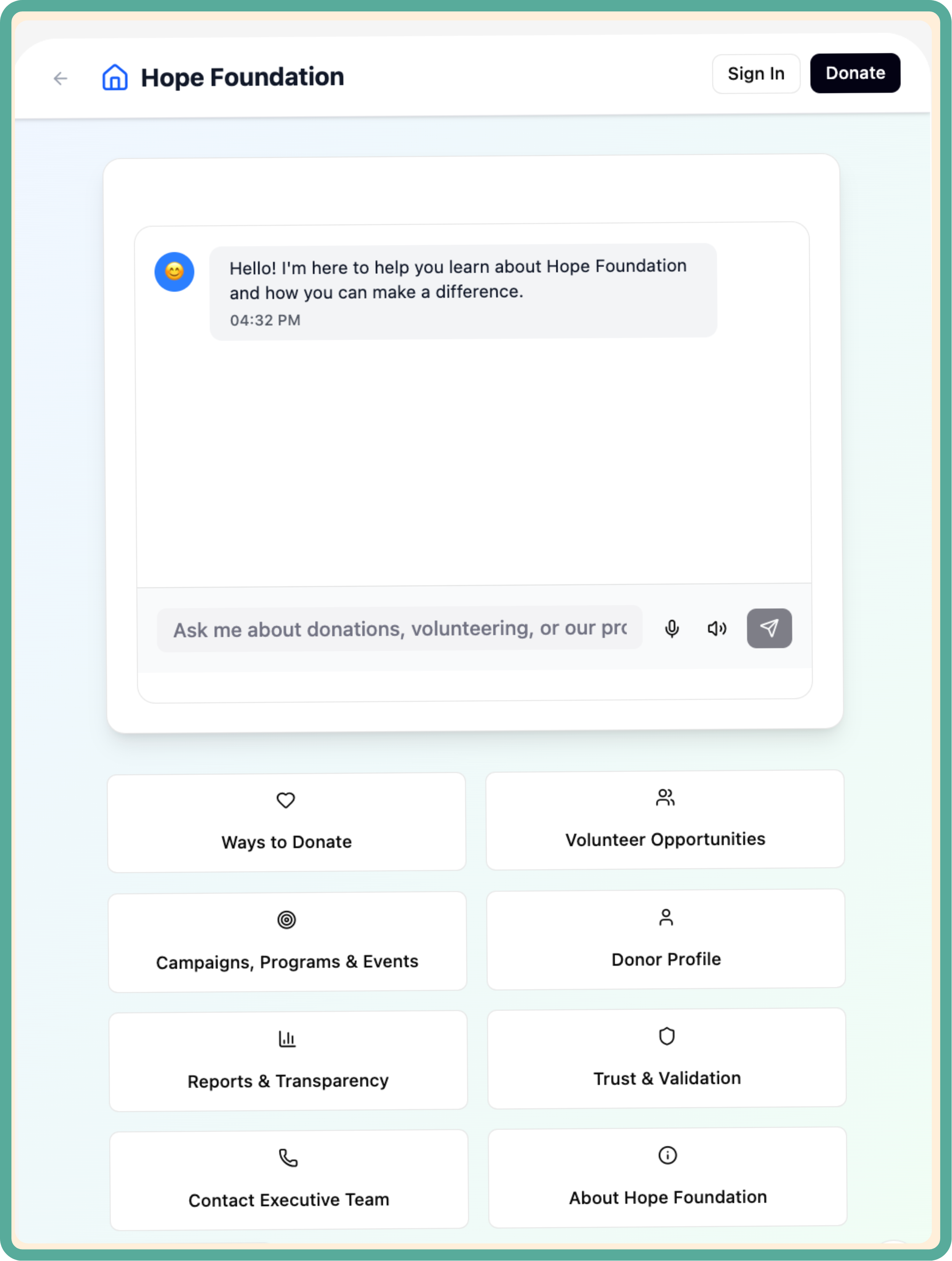}
    \end{minipage}
    \hfill
    \begin{minipage}{0.35\textwidth}
        \caption{The Local Agent page of AMINA, featuring the chat interface and a set of simplified action buttons tailored to donor needs.}
        \label{fig:AMINA-b.png}
    \end{minipage}
\end{figure}

\begin{figure}[h]
    \centering

    \begin{minipage}{0.6\textwidth}
        \includegraphics[width=\linewidth]{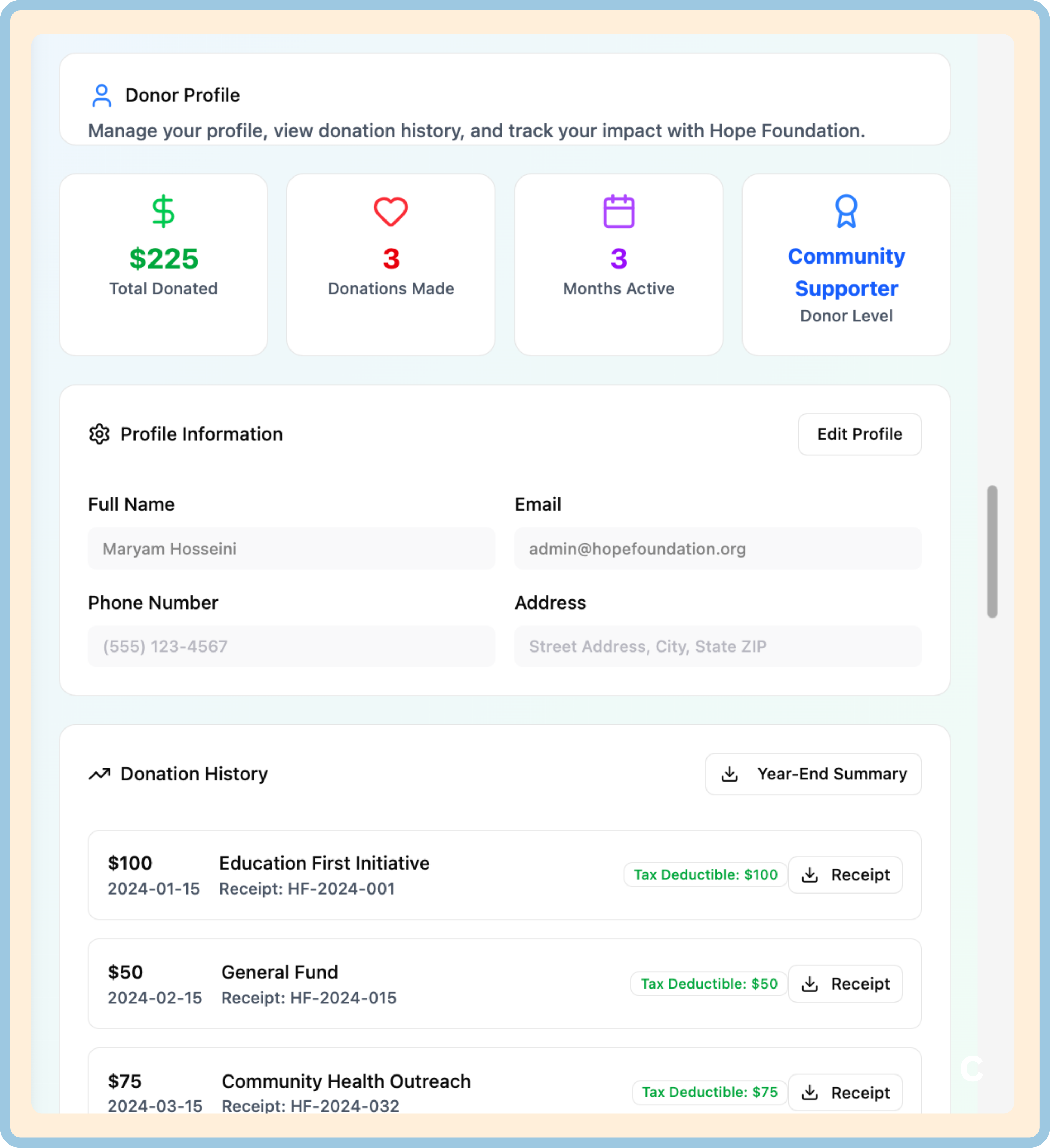}
    \end{minipage}
    \hfill
    \begin{minipage}{0.35\textwidth}
        \caption{The donor profile, indicating an overview of impacts, profile information, and transaction history.}
        \label{fig:AMINA-c.png}
    \end{minipage}
\end{figure}

\begin{figure}[h]
    \centering

    \begin{minipage}{0.6\textwidth}
        \includegraphics[width=\linewidth]{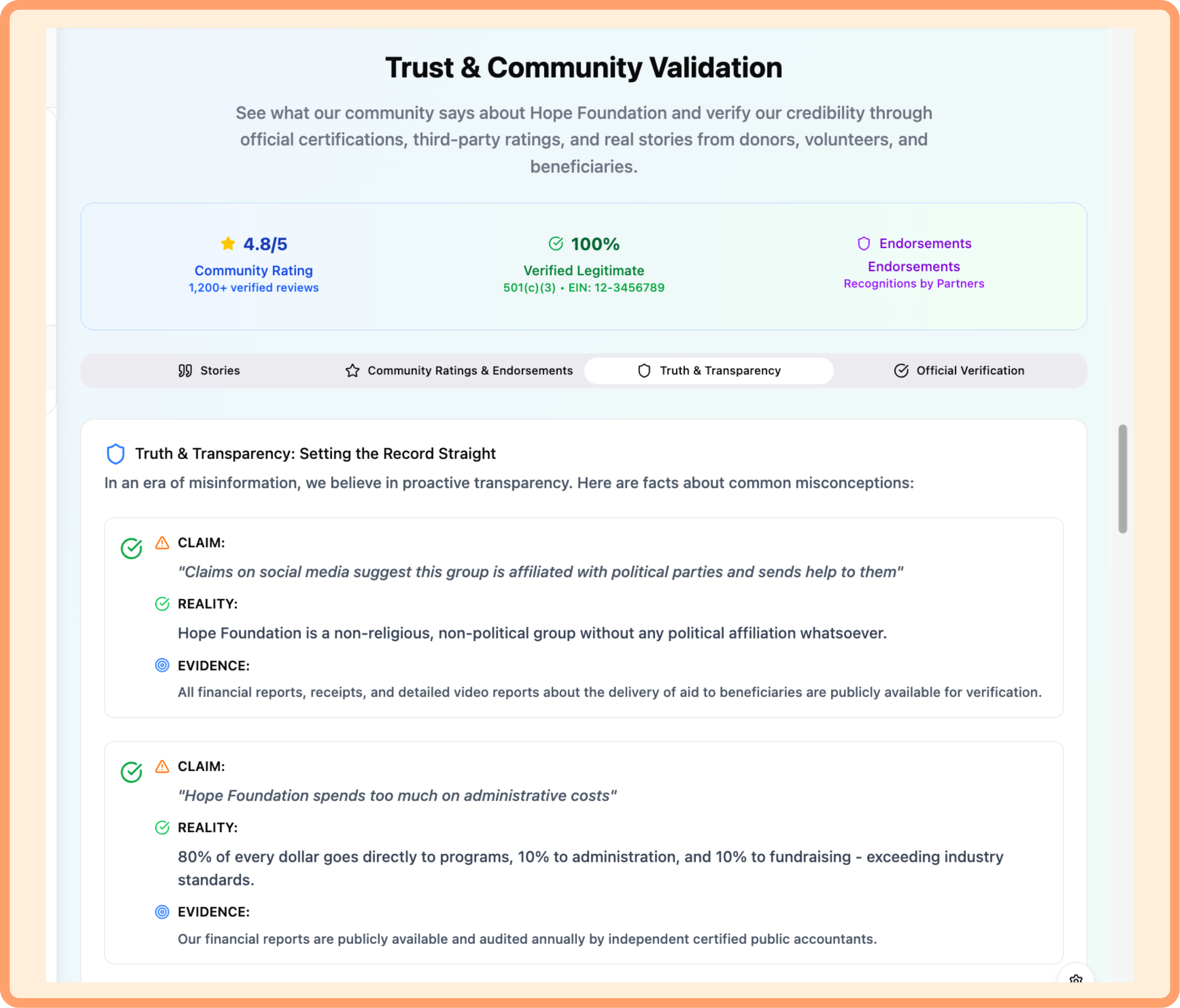}
    \end{minipage}
    \hfill
    \begin{minipage}{0.35\textwidth}
        \caption{Debunking misinformation tab, indicating design artifacts that encourage nonprofits for proactive debunking of misinformation (DG3) by providing "fact-checking" buttons (in green, on the left), fields to include viral misinformation claims, true realities, and clickable sections to provide evidence.}
        \label{fig:AMINA-d.png}
    \end{minipage}
\end{figure}

\begin{figure}[h]
    \centering

    \begin{minipage}{0.6\textwidth}
        \includegraphics[width=\linewidth]{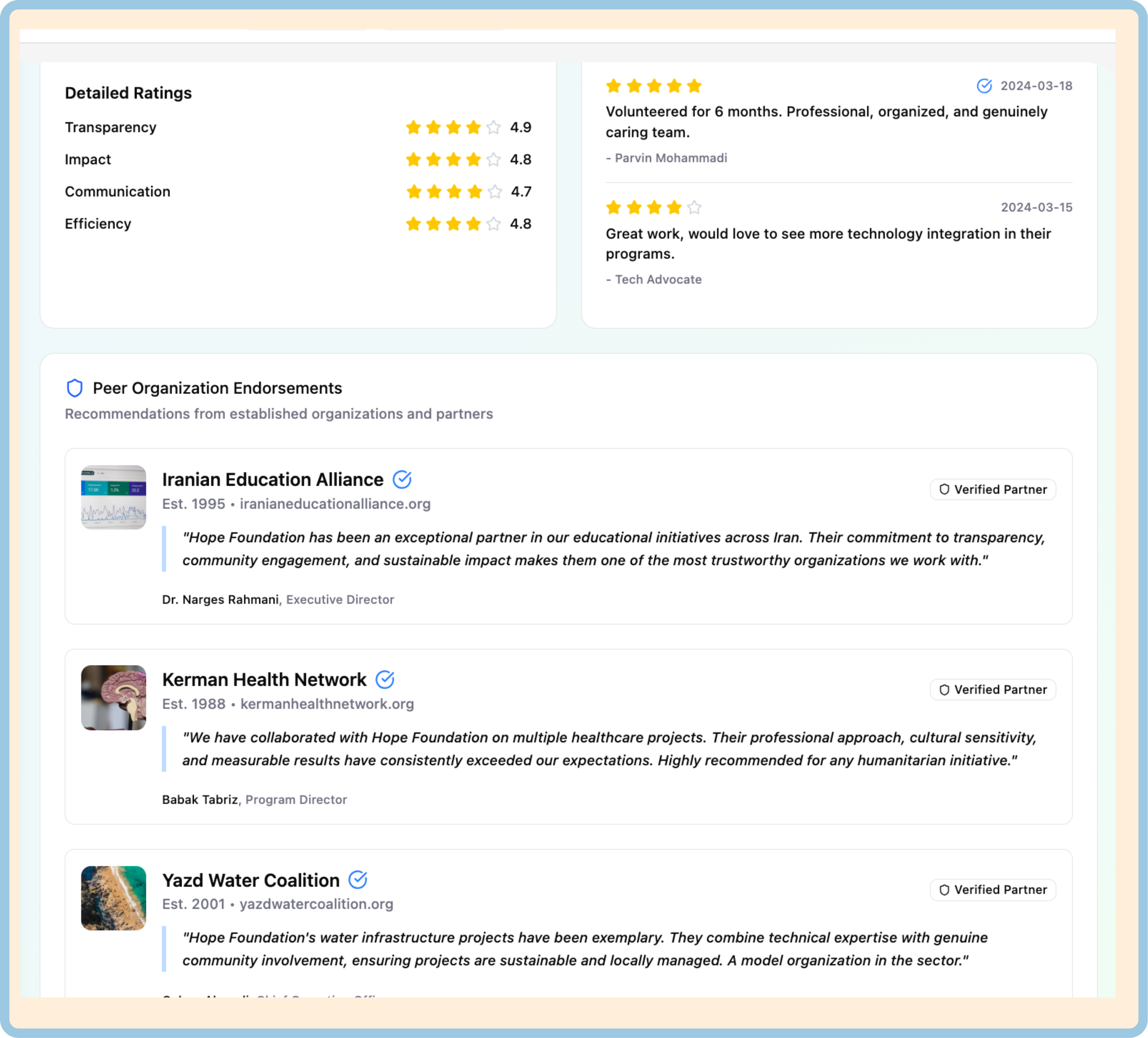}
    \end{minipage}
    \hfill
    \begin{minipage}{0.3\textwidth}
        \caption{The "multi-criteria rating system and peer endorsement tab" that facilitates community evaluation of nonprofits.}
        \label{fig:AMINA-e.png}
    \end{minipage}
\end{figure}

\begin{figure}[h]
    \centering

    \begin{minipage}{0.6\textwidth}
        \includegraphics[width=\linewidth]{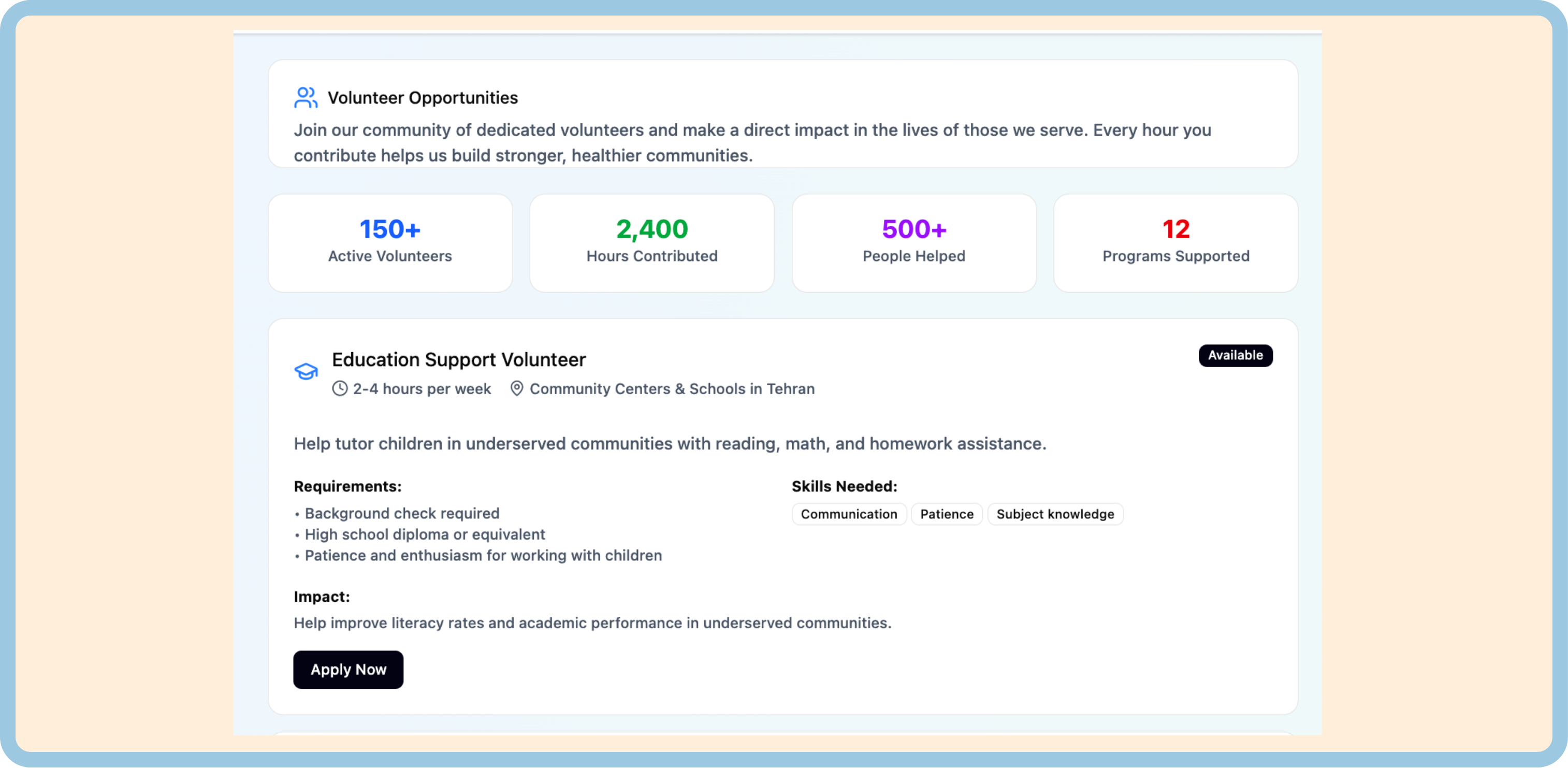}
    \end{minipage}
    \hfill
    \begin{minipage}{0.3\textwidth}
        \caption{The volunteer opportunities page, indicating a sample volunteer hiring form.}
        \label{fig:AMINA-f.png}
    \end{minipage}
\end{figure}

\begin{figure}[h]
    \centering

    \begin{minipage}{0.5\textwidth}
        \includegraphics[width=\linewidth]{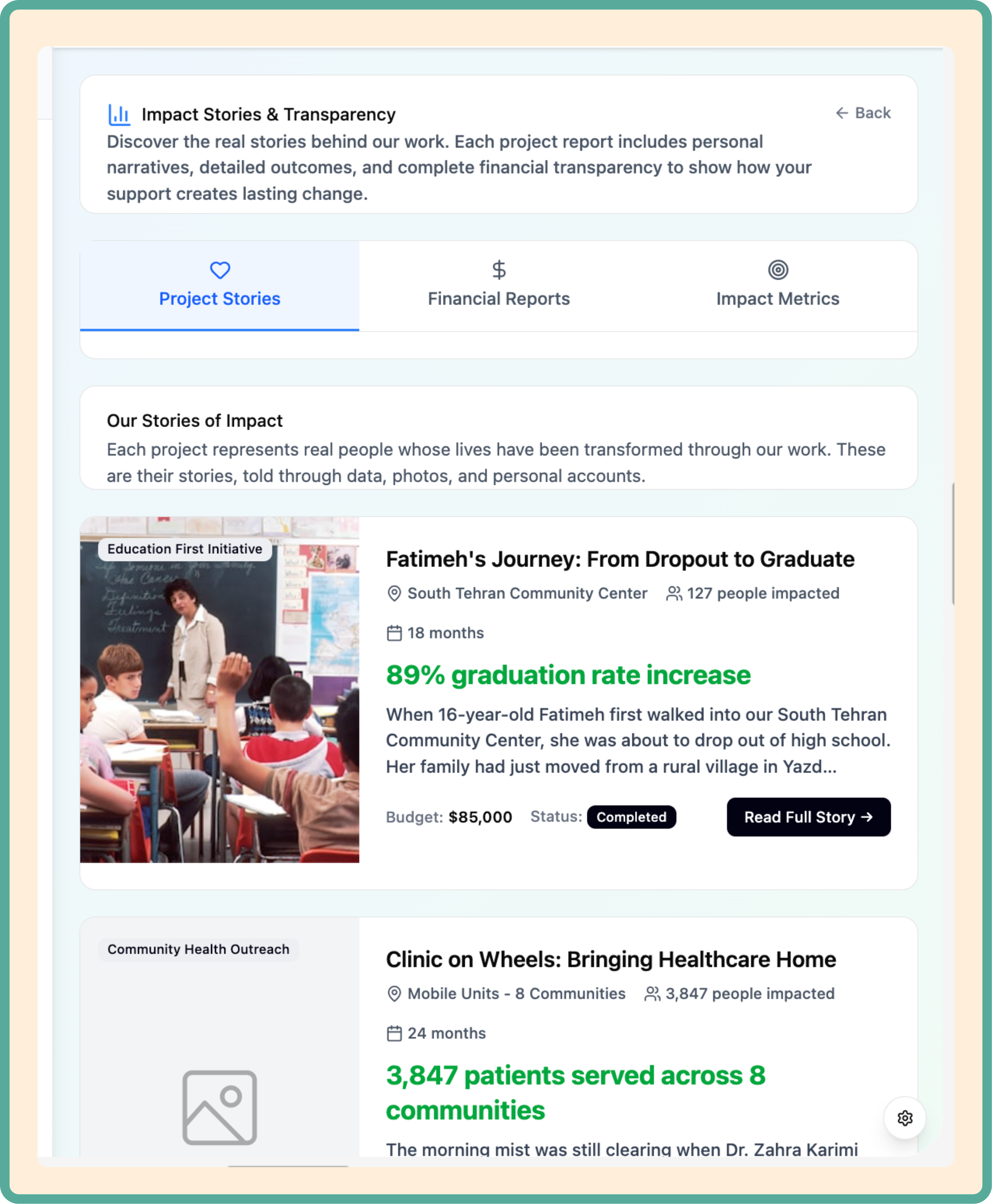}
    \end{minipage}
    \hfill
    \begin{minipage}{0.35\textwidth}
        \caption{The reports and transparency page, here showing the qualitative stories tab.}
        \label{fig:AMINA-g.png}
    \end{minipage}
\end{figure}

\begin{figure}[h]
    \centering

    \begin{minipage}{0.5\textwidth}
        \includegraphics[width=\linewidth]{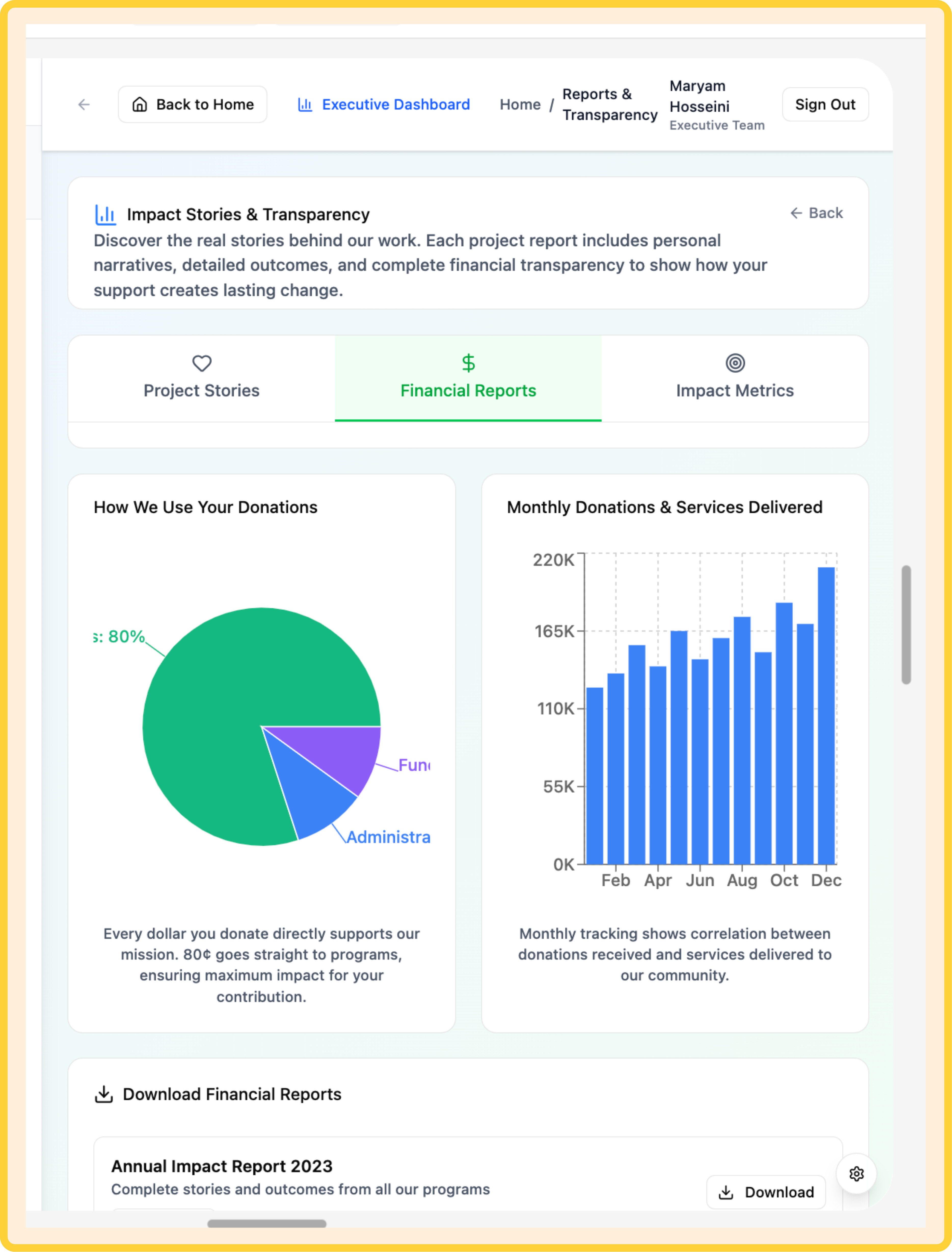}
    \end{minipage}
    \hfill
    \begin{minipage}{0.35\textwidth}
        \caption{The reports and transparency page, here showing the quantitative reports tab.}
        \label{fig:AMINA-h.png}
    \end{minipage}
\end{figure}

\begin{figure}[h]
    \centering

    \begin{minipage}{0.55\textwidth}
        \includegraphics[width=\linewidth]{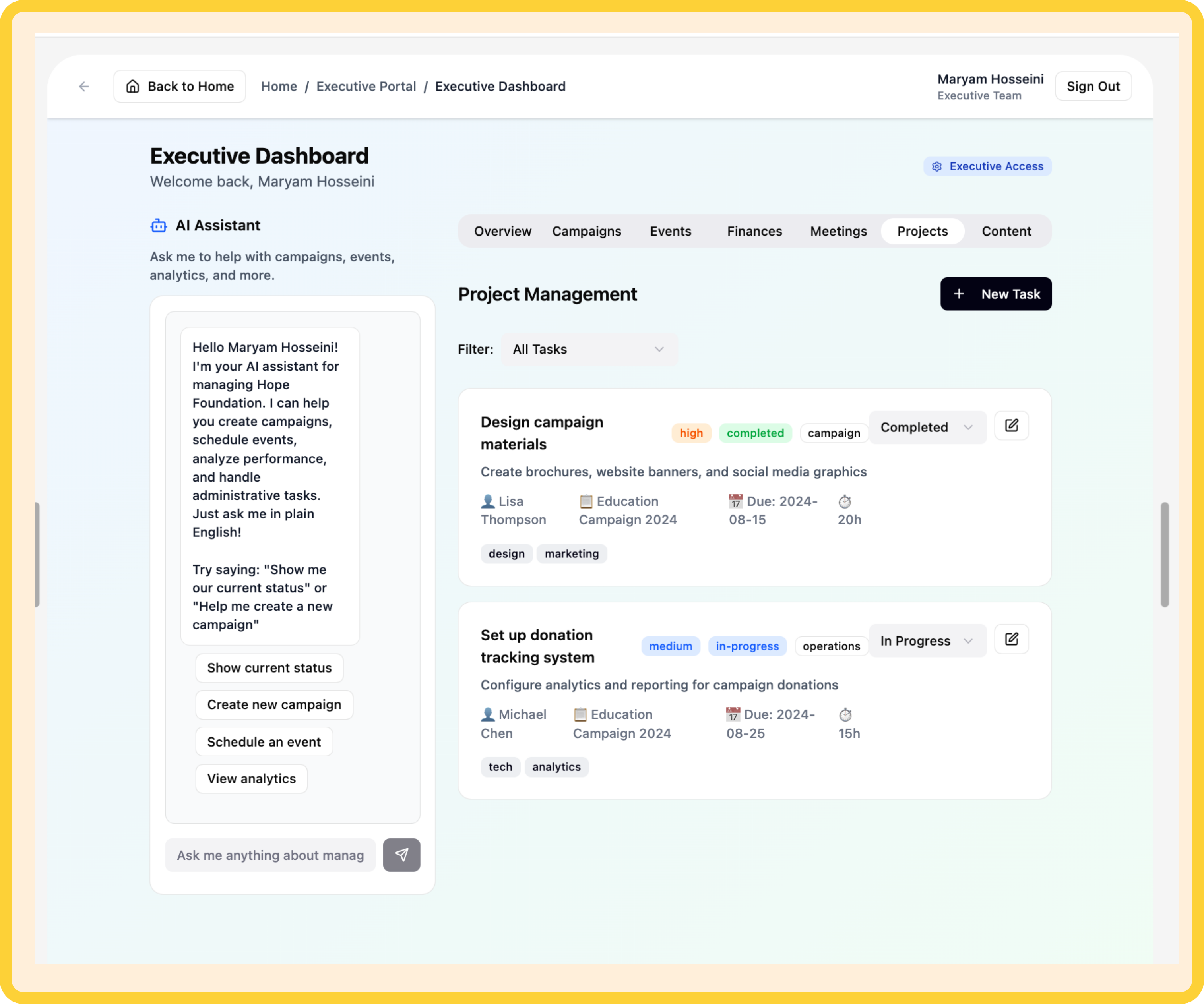}
    \end{minipage}
    \hfill
    \begin{minipage}{0.35\textwidth}
        \caption{The executive dashboard, with the project management tab activated.}
        \label{fig:AMINA-i.png}
    \end{minipage}
\end{figure}

\begin{figure}[h]
    \centering

    \begin{minipage}{0.55\textwidth}
        \includegraphics[width=\linewidth]{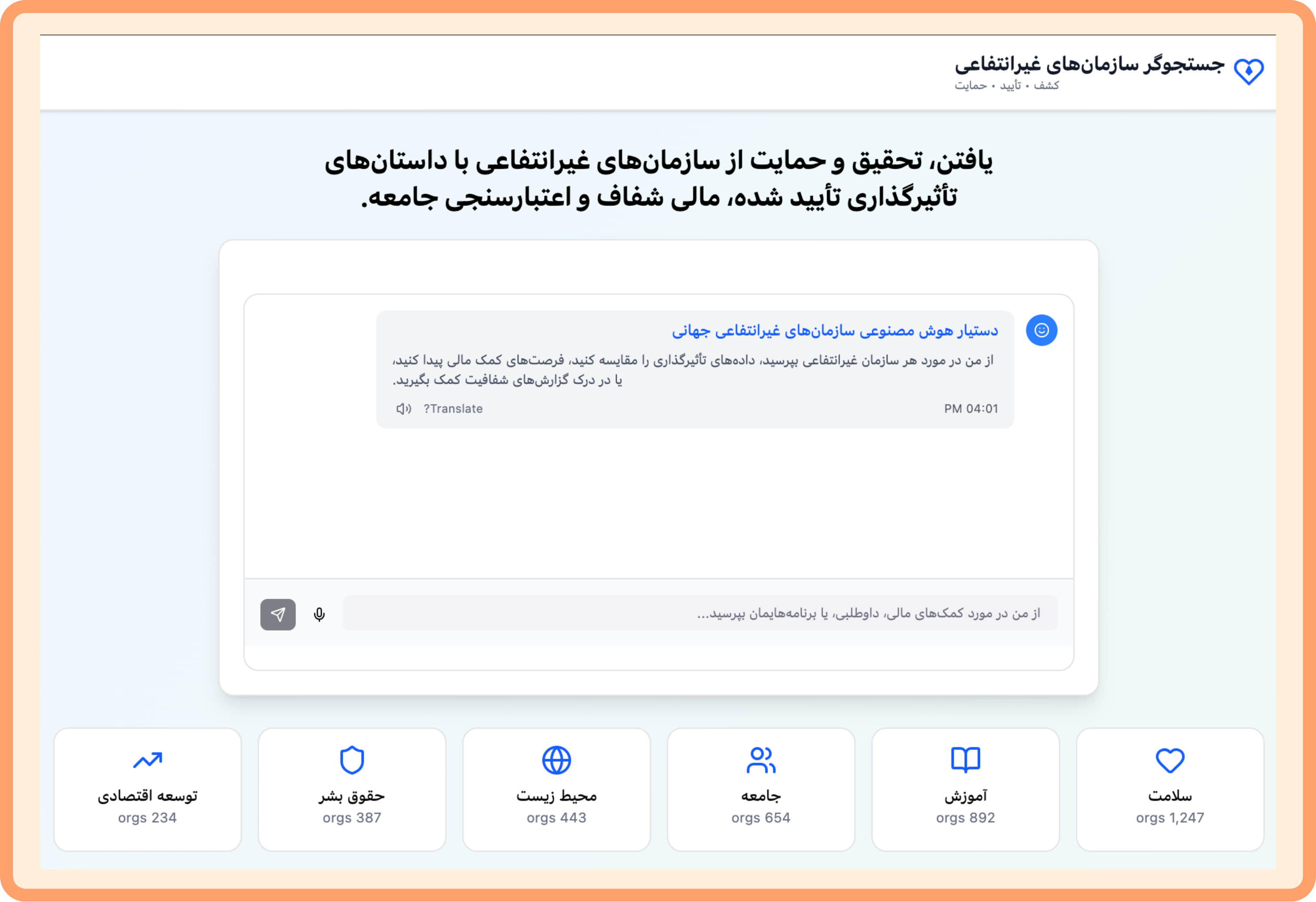}
    \end{minipage}
    \hfill
    \begin{minipage}{0.35\textwidth}
        \caption{An example of the platform’s Farsi translation.}
        \label{fig:AMINA-j.png}
    \end{minipage}
\end{figure}

\end{document}